\documentclass[useAMS,usenatbib,usegraphicx]{mn2e}
\usepackage{rotating}
\usepackage{txfonts}

\def\lae{\mathrel{<\kern-1.0em\lower0.9ex\hbox{$\sim$}}}
\def\gae{\mathrel{>\kern-1.0em\lower0.9ex\hbox{$\sim$}}}

\title[The evolution of early-types since $z\sim2$]
{The population of early-type galaxies at $1<z<2$ - New clues on their
formation and evolution}
\author[P. Saracco et al.]{P. Saracco$^{1}$\thanks{E-mail: 
paolo.saracco@brera.inaf.it}, M. Longhetti$^{1}$,  S. Andreon$^{1}$\\ 
$^{1}$INAF - Osservatorio Astronomico di Brera, Via Brera 28, 20121 Milano
}
\begin{document}

\setcounter{table}{1}

\date{Accepted 2008 October 13. Received 2008 Ocober 3; in original form 2008
August 18.}
\pagerange{\pageref{firstpage}--\pageref{lastpage}} \pubyear{2008}
\maketitle
\label{firstpage}
\begin{abstract}
We present the morphological analysis based on HST-NICMOS 
observations in the F160W filter ($\lambda\simeq1.6$ $\mu$m) 
of a sample of 32 early-type galaxies (ETGs)
at $1<z<2$ with spectroscopic confirmation of their redshift and spectral type.
The 32 ETGs at $\langle z\rangle\sim1.5$ are placed on the 
[$\langle\mu\rangle_e$, R$_e$]
plane according to a Kormendy relation with the same slope of the local one
but with a different zero-point which accounts for the evolution they undergo 
from $z\sim1.5-2$ to $z=0$.
The best fitting of their SED shows that ETGs at $1<z<2$ are composed of two 
distinct populations, an older population (oETGs) and a younger 
population (yETGs) whose mean ages differ by about 1.5-2 Gyr.
Young ETGs are not denser than local ones since they follow the size-mass 
relation of local ETGs and luminosity evolution brings them  onto the local 
Kormendy and size-luminosity relations without the need of size evolution.
Old ETGs do not follow the size-mass relation of local ETGs and luminosity
evolution does not account for the discrepancy they show with respect to
the local size-luminosity and Kormendy relations.
An increase  of their R$_e$  by a factor 2.5-3 (a density decrease
by a factor 15-30) from $z\sim1.5-2$ to $z\sim0$
is required to bring these galaxies onto the local scaling relations.
The different properties  and the different
behaviour shown by the two populations with respect to the scaling relations 
imply different formation and evolution scenarios.
The older population of ETGs must have formed at higher-z in a sort of 
dissipative gas-rich collapse ables to produce remnants which at 
$z\sim2$ are old and compact, scenario which can be fitted qualitatively 
by some recent hydrodynamic simulations of gas-rich mergers. 
Given the typical time scale of merging and the old age of their stellar 
population, oETGs should exist as they are up to $z\gae3-3.5$.
The size evolution they must experience from $z\sim2$ to $z\sim0$ must leave 
unchanged their mass to not exceed the local number of high-mass 
($\mathcal{M}_*>5\cdot10^{11}$ M$_\odot$) ETGs.
Thus, major merging cannot fit this requirement. 
Satellite merging, close encounters and interactions can help at least
qualitatively in solving this problem.
The younger population of ETGs can be formed later through subsequent episodes
of merging which increased progressively their size and assembled their 
mass down to $z\sim2$.
At $z<2$ they evolve purely in luminosity since episodes of major merging would
bring them far from the local scaling relations.

paolo.saracco@brera.inaf.it
\end{abstract}

\section{Introduction}
The formation and the evolution of early-type galaxies (ETGs, elliptical
and bulge-dominated galaxies) occupy an important position among the 
challenges of the observational cosmology.
At least $\sim70\%$ of the stellar mass in the local universe is 
locked into ETGs.
For this reason, the understanding of their build-up and growth is fundamental
to trace the galaxy mass assembly in the Universe.
Their homogeneous properties, e.g. colors and  scaling relations,
make them an excellent 
probe to investigate the history of the stellar mass assembly of
galaxies over cosmic times (see e.g. Renzini et al. 2006 and references 
therein).
Most of the recent studies based on samples of ETGs at $z<1$  agree with 
considering completed their build up at $z\sim0.8$.
This statement is supported by the results
on the evolution of the stellar  mass function of galaxies which do not show 
any deficit of high-mass galaxies up to $z=0.8-1$ 
(e.g. Fontana et al. 2004; Pozzetti et al. 2007),
by the observed evolution of the bright end of the luminosity function of 
galaxies consistent with the pure luminosity evolution expected for 
early-types (e.g. Drory et al. 2005; Saracco et al. 2006; Zucca et al. 2006;
Caputi et al. 2007; Cirasuolo et al. 2007; 2008), by 
the observed number density of 
ETGs at $z\le1$ consistent with the one at $z=0$  
(e.g. Saracco et al. 2005;  Cimatti et al. 2006; Conselice et al. 2007)
and by the evolution of the size-mass and size-luminosity relations
compatible with a passive luminosity evolution 
(e.g. McIntosh et al. 2005).
If on one hand the agreement among these results  provide a clear view 
of the status and of the evolution of ETGs at $z<1$, they do not add 
stringent constraints on the mechanism(s) with which ETGs assemble 
their mass, pushing at higher $z$ the redshift range of interest.

Several studies, indeed, suggest that both the observational and the 
theoretical efforts aimed at constraining  the build up of the stellar mass
of ETGs and the shaping of their morphology 
should be focused  at $1<z<2$, the redshift range for which the 
strongest evolution is expected 
(e.g. Glazebrook et al. 2004; Arnouts et al. 2007).
The picture at this redshift is far from being as clear and
homogeneous as at $z<1$ because of the difficulties in catching
ETGs at high redshift.
Indeed, to study the population of ETGs at $1<z<2$, a preliminary
but not so obvious step has to be overcome: the identification of 
suitable samples of ETGs  with secure redshift determination, 
spectral classification and confirmed morphological signatures 
similar to those of the local ETGs. 
In fact, up to now, only few 
samples of spectroscopically identified ETGs at $z>1$ have been collected, 
with no more than a tenth of galaxies morphologically confirmed each:  
the sample of McCarthy et al. (2004) resulting from the GDDS
(Abraham et al. 2004)  contains 10 galaxies 
at $z\sim1.6$; the sample of Longhetti et al. (2005) derived from
the TESIS (Saracco et al. 2003) contains 
10 galaxies at $z\sim1.4$ and the sample of Cimatti et al. (2008) 
resulting  from the K20 (Cimatti et al. 2002) and the GMASS  
surveys contains 13 galaxies at $z\sim 1.6$.
On the basis of the analysis of these few samples, 
it is well ascertained that ETGs at $z \sim1-2$ contain stellar 
populations formed at $z>3$ in an intense and short lived starburst 
(Longhetti et al. 2005; Kong et al. 2006; Farrah et al. 2006; Cimatti et al.
2004; McGrath et al. 2007).
This information provides a strong constraint on the star formation
history of the stellar population of ETGs, that is the epoch at which the 
stars they host formed.
On the other hand, this does not constrain how ETGs grew up:
 what are the time scale and the mechanism(s)
characterizing the growing and the shaping of ETGs ?

Recently, evidence for higher compactness of the ETGs at $z>1$ with respect 
to the local ones have come out.
Daddi et al. (2005) show that a large fraction of their ETGs
have smaller sizes (effective radii R$_e\sim1$ kpc) than local 
ETGs of comparable stellar mass, possibly implying higher
stellar densities even if the presence
of AGN in some of them could justify the compactness.
However, other studies find similar results confirming the apparent smaller
sizes of high-z ETGs if compared to the local ETGs with comparable
stellar mass (e.g. Di Serego Alighieri et al. 2005; Trujillo et al. 2006a; 
Cassata et al. 2005).
These results are based on HST optical observations sampling the blue and UV 
rest-frame emission of the galaxies and/or on seeing limited 
ground-based observations,  characteristics which could affect the estimate 
of the effective radius of high-z early-type galaxies.
Moreover, the above results have been obtained by comparing galaxies at 
different redshift having, in principle, the same stellar mass.
However, the stellar mass estimate depends on the spectrophotometric 
models used to fit the data and on the different model parameters.  
More recently, Longhetti et al. (2007) have studied the Kormendy relation
for a sample of ETGs at $z\sim1.5$ using HST-NICMOS observations.
They show that these ETGs are at least 2 times more compact than those in 
the local Universe showing that this apparent high compactness is real
and not dependent on the wavelength of observation 
(see also McGrath et al. 2008 and Buitrago et al. 2008 for similar
recent results).
Some of these works are based on the small samples quoted
above made of a tenth of galaxies  spanning a narrow
range in luminosity and stellar mass or on samples of candidates ETGs
with no confirmation of their redshift and spectral type.

In an attempt to provide  new and stronger constraints on the formation
of  compact/dens high-z ETGs we have studied the main scaling relations
(the Kormendy, the size-luminosity and the size-mass relations) for a 
new sample of 32 ETGs at $1<z<2$ with spectroscopic confirmation  of 
their redshift and spectral type.
The morphological analysis for the whole sample is based on HST-NICMOS 
imaging in the F160W filter ($\lambda\sim1.6$ $\mu$m) which samples the 
rest-frame R-band at the redshift of the galaxies.
The sample spans 3 magnitudes in absolute magnitude and more than
two orders of magnitude in stellar mass.
This paper presents the analysis and the results we obtained from 
the study of the scaling relations and it is organized as follows.
Section 2 is a presentation of our sample describing the criteria used 
to construct the sample and the data we have at hand.
Section 3 describes the methodology used to determine the main 
physical properties (effective radius, surface brightness, absolute magnitude,
stellar mass and age) of the 32 ETGs.
Section 4 presents the Kormendy relation while Section 5 presents a discussion
of the results obtained in Sec. 4.
Section 6 shows the size-luminosity and the size-mass relations while Section 7
places the results in the context of the galaxy formation and evolution 
scenarios. Section 8 is a summary.

Throughout this paper we use a standard cosmology with
$H_0=70$ Km s$^{-1}$ Mpc$^{-1}$, $\Omega_m=0.3$ and $\Omega_\Lambda=0.7$.
All the magnitudes are in the Vega system, unless otherwise specified.

\section{Sample selection and HST-NICMOS imaging}
The sample of  early-type galaxies  we constructed is composed of 
32 galaxies at $1<z<2$ selected from different samples and surveys 
on the basis of their spectroscopic and morphological
classification.
We restricted our selection to those galaxies having both
$i)$ deep HST-NICMOS observations in the F160W filter  
($\lambda\sim1.6$ $\mu$m) sampling the rest-frame continuum 
 $0.55<\lambda_{rest}<0.85$  $\mu$m at  $1<z<2$  and $ii)$ 
 spectroscopic confirmation of their redshift and spectral type.
On the basis of these criteria we were able to collect a sample of
32 ETGs. 
The samples from which they have been extracted are the following:

\noindent
- 10 ETGs  at $1.4<z<1.9$ have been selected from  
the Galaxy Deep-Deep Survey sample (GDDS, Abraham et al. 2004;
McCarthy et al. 2004).
According to the spectral classification described in
Abraham et al., we selected  those galaxies  having Class=001
(8 galaxies), i.e. pure signatures of an evolved stellar population, 
and two galaxies having signatures of a young population superimposed
to the older one (Class=101 and 011). These galaxies are listed
in Tab. 1 as SA\# and have been recently studied also 
by Damjanov et al. (2008).

\noindent
- 6 ETGs at $z\simeq1.27$ have been selected from the sample of 
Stanford et al. (1997) and belong to the cluster RDCS 0848+4453 
in the Linx field.
Their spectra show absorption features (Ca II H+K, Mg I and Mg II)
and spectral break (B2900, D4000) similar to the present epoch 
ellipticals (Stanford et al. 2007; Van Dokkum et al. 2003).
These galaxies, listed in Tab. 1 as CIG\#, have been previously 
studied also by Moriondo et al. (2000).

\noindent
- 3 ETGs at $1<z<1.8$ have been selected from the sample of 
Stanford et al. (2004) in the Hubble Deep Field-North (HDF-N)
 according to their
spectral type ST$<0.1$ characterizing an old and passive population
of stars and on their morphology.  They are listed as HDF\_\#.

\noindent
- 2 ETGs (HUDF\_\#) at $z\simeq1.4$ and at $z\simeq1.9$ respectively  
have been selected from the sample of Cimatti et al. 
(2008; see also Daddi et al. 2005) in 
the Hubble Ultra Deep Field (HUDF). 
They are classified as early-types both on the basis
of their  morphology and on their spectral features.

\noindent
- the ETG 53W091 has been taken from the paper of Dunlop et al. (1996;
see also Spinrad et al. 1996).
The spectrum of this galaxy is characterized by absorption features
typical of an old stellar population and its light profile is bulge 
dominated (Waddington et al. 2002). 

\noindent
- The remaining 10 ETGs  (S2F\# and S7F\#) come from our own sample of 
ETGs spectroscopically classified at $1.2<z<1.7$ in the framework of the 
TESIS project (Saracco et al. 2003, 2005).
The study of their spectro-photometric properties and of their morphology  
based on multiwavelength data and HST-NICMOS observations
are described in previous works (Longhetti et al. 2005, 2007).

The whole sample of 32 ETGs is listed in Table 1 where
we also report for each galaxy the photometry  in different bands.
All the magnitudes, with the exception of the F160W-band magnitude, 
are taken from the literature as quoted in the Column 4 of the table. 
The magnitude in the F160W filter is the SExtractor MAG\_BEST magnitude
(Bertin and Arnouts 1997) that we estimated from the HST-NICMOS images 
we retrieved from the HST archive. 
All the magnitudes are in the Vega system.
Given the different samples the galaxies have been extracted from, 
the wavelength 
coverage is not the same for all the galaxies as well as the filters used.
We did not convert the magnitudes derived in slightly different filters 
from the different surveys to a common filter system since we would
have introduced large and possibly systematic errors.
We have preferred to keep the original magnitudes as given by the authors 
and to use the appropriate set of response filter functions in our analysis.
We describe the multiwavelength coverage and the filters 
used in the various surveys in Appendix A.

The median redshift of the sample thus collected is $z_{med}=1.45$.
HST-NICMOS images with the NIC2 (0.075 arcsec/pix) camera 
are available for the 10 galaxies of our sample and
for the galaxy 53W091, i.e. for $\sim30\%$ of the sample. 
For the remaining galaxies the available images are
based on NIC3 (0.2 arcsec/pix)  camera.
The NIC3 images relevant to the two galaxies in the HUDF were 
drizzled to 0.09 arcsec/pixel.
{ The 1-$\sigma$ limiting surface brightness $\mu_{lim}$ 
of the different NICMOS images is reported in Tab. 1.}
The NICMOS  mosaics for the whole sample of ETGs can be 
retrieved at the web page \texttt{http://www.brera.inaf.it/utenti/saracco/}.

\begin{figure*}
\vskip -2truecm
\hskip -2.0truecm
\includegraphics[width=20.cm, height=27cm]{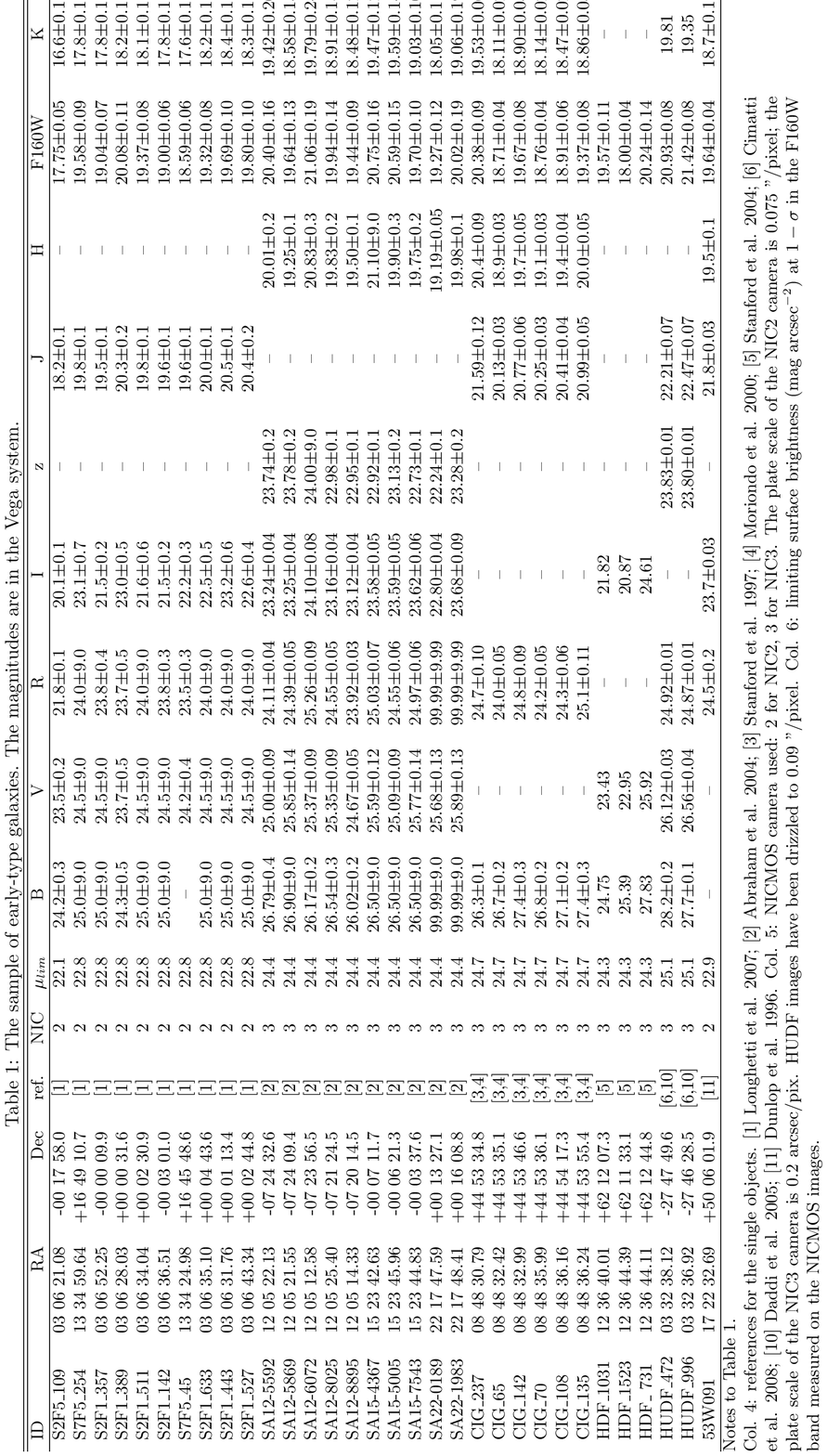}
\end{figure*}

\section{Physical parameters of early-type galaxies}

\subsection{Morphological parameters}
We derived the effective radius r$_e$ [arcsec] and the mean surface brightness
(SB) $\langle\mu\rangle_e$ [mag/arcsec$^2$] within  r$_e$ of our galaxies 
from the NICMOS images by fitting a S\'ersic profile (S\'ersic 1968) to the 
observed light profiles. 
The analytic expression of the adopted profiles is
\begin{equation}
I(r)=I_{e} exp\{-b_{n}[(r/r_{e})^{1/n}-1]\}
\end{equation}
where $n=4$ and $n=1$ values define the de Vaucouleurs (de Vaucouleurs 1948) 
and  the exponential (disk) profiles respectively. 
We used \texttt{Galfit} software (v. 2.0.3; 
Peng et al. 2002) to perform the fitting to the observed profiles.
The bi-dimensional S\'ersic model has been convolved with the 
Point Spread Function (PSF) of the NIC2 and the NIC3 
cameras modeled by means of the 
\texttt{Tiny Tim}\footnote{www.stsci.edu/software/tinytim}
(v. 6.3) software package (Krist 1995; Krist J. \& Hook R. 2004). 
The fitting provided us with the semi-major axis $a_e$ of the projected
elliptical isophote containing half of the total light and with
the axial ratio $b/a$.
We thus derived the circularized effective radius $r_e=a_e\sqrt{b/a}$.

To assess the robustness and the accuracy of our estimate of the 
effective radius of galaxies, 
we applied the same fitting procedure to a set of simulated galaxies
inserted in the real background. 
The simulations follow those described in details in Longhetti et al. (2007)
for the NIC2 images.
Here, we summarize the main features of the procedure followed to
obtain the simulated observations.
We  generated with \texttt{Galfit} a set of 100 galaxies 
described by a  de Vaucouleurs profile  with  axial ratio $b/a$ 
and position angle  PA randomly assigned in the ranges $0.4<b/a<1$ 
and 0$<$PA$<$180 deg respectively.
Magnitudes in the F160W filter were  assigned randomly  in the range
$18<F160W<21$.
{ Effective radii $r_{e,in}$ were assigned randomly  in the ranges 
$0.2<r_{e,in}<0.5$ arcsec (corresponding to 1.5-4 kpc at $z\simeq1$)
for the NIC2 images. 
In order to verify the absence of a bias against the detection of small 
effective radii ($r_{e,in}<0.2$ arcsec, R$_e<1.5$ Kpc) in the fitting
of galaxies in the NIC3 images characterized by a pixel scale of 
0.2 arcsec/pix, we simulated here also a set of galaxies with 
$0.05<r_{e,in}<0.2$ arcsec.
The simulated galaxies have been convolved with the NICMOS PSFs  
and then embedded in the real NIC2 and NIC3 images.
We used the simulations described in Longhetti et al. (2007) 
for the NIC2 galaxies and the images relevant to the sample
of McCarthy et al. (2005) to simulate the NIC3 galaxies.}
We then  fit the simulated galaxies with the S\'ersic profile
and studied the behaviour of the resulting $r_{e,fit}$  checking our
ability in recovering  the input value $r_{e,in}$.
In Figure 1 we plot the values of the effective radius $r_{e,fit}$  
 versus $r_{e,in}$ in the case of NIC2 and NIC3 images. 
It can be seen that  in the NIC2 images the effective
radius of the galaxies is slightly underestimated by 
$\Delta r_e^{NIC2}\simeq0.07$ arcsec on average, while in the NIC3
images the mean underestimate is  
$\Delta r_e^{NIC3}\simeq0.03$ arcsec.
The rms measured is 0.04 arcsec in the NIC3 images and 0.02 arcsec in the NIC2
images and  are much larger than the formal
fitting error on the effective radius. 
\begin{figure}
\begin{center}
\includegraphics[width=9cm]{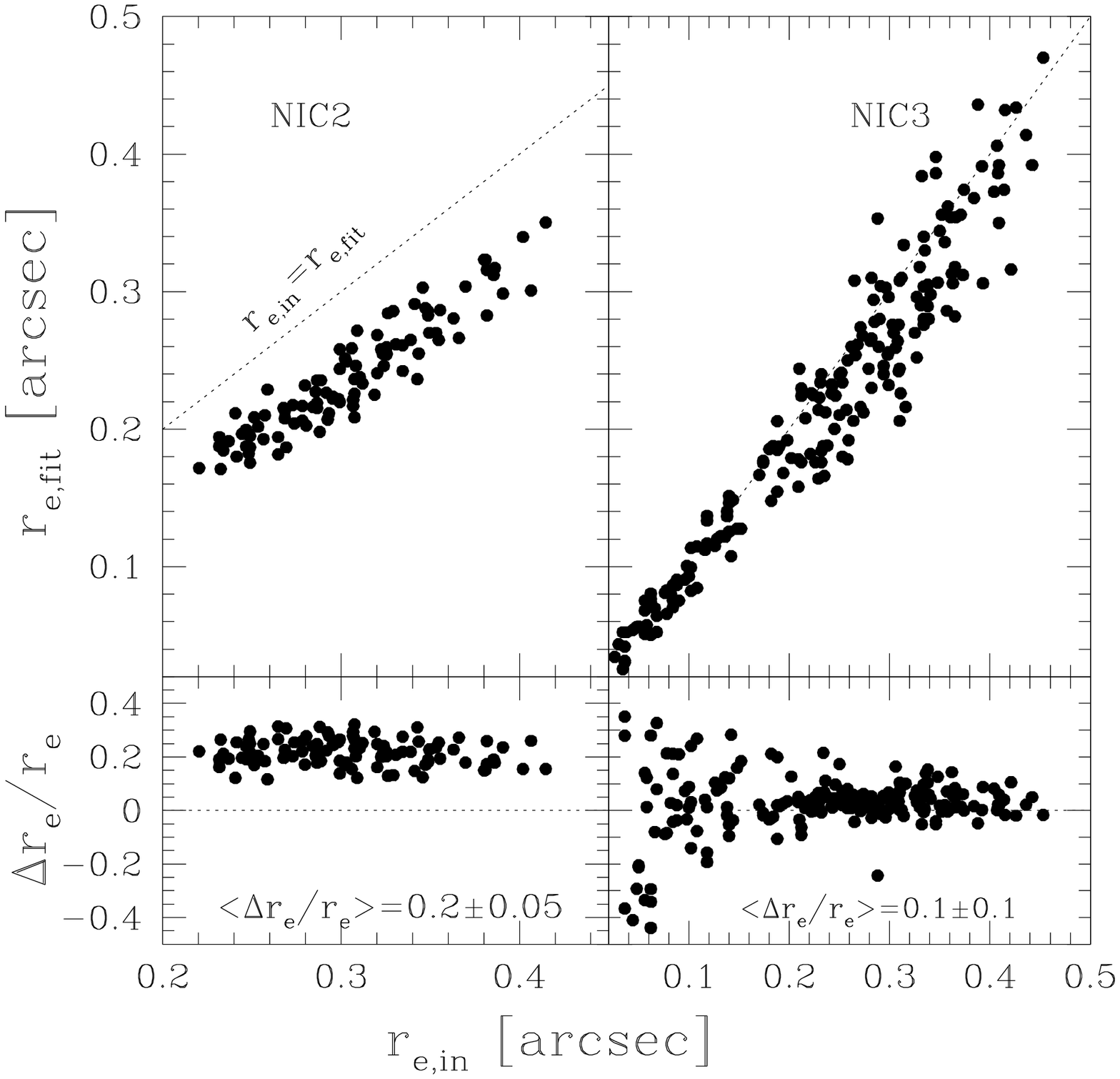}
\caption{Comparison between the effective radius of the simulated galaxies
(r$_{e,in}$) and the effective radius r$_{e,fit}$ obtained through the 
fitting with the S\'ersic profile for the NIC2 (left panel) and the NIC3 (right
panel) images.}
\end{center}
\end{figure}
Thus our fitting tends to slightly underestimate 
the effective radius $r_{e}$.
{ Fig. 1 shows also that the pixel scale of NIC3 does not
represent a limit in the detection of small effective radii and thus that
no bias is present in our analysis against very small galaxies.} 
The small underestimate of $r_e$ has been taken into account in the following
analysis and in the derivation of the mean surface brightnesses by adding 
the mean offset $\Delta r_e$ quoted above to the best fitting values $r_e$.
In Tab. 2 we report the morphological parameters $r_e$ [arcsec] and R$_e$ [Kpc] 
 derived from the fitting to the profile  of our galaxies and the mean SB
in the F160W band
\begin{equation}
\langle\mu\rangle_e^{F160W}=F160W_{tot}+5log(r_e)+2.5log(2\pi)
\end{equation}
{ where F160W$_{tot}$ is the total magnitude in this filter  
derived by \texttt{galfit} and reported in Tab. 2}.

The morphology for some of them, namely the CIG\# galaxies, 
the HDF\# galaxies and  53W091, had been already derived in the rest-frame 
R-band from Moriondo et al. (2000), Stanford et al. (2004) and 
Waddington et al (2002) respectively.
Our new estimates are in good agreement with their estimates in spite of
the different method used  to fit the profiles.
For four galaxies of our sample, namely HUDF\_472, HUDF\_996, HDF\_1031 
and HDF\_1523, the estimate of the effective radius derived from HST images
in the optical bands (F814W and F850W filters) sampling the rest-frame 
wavelength $\lambda_{rest}<4000$ \AA, is also available from the 
literature.
In the case of the two galaxies in the HDF at $z\sim1$ the estimates of 
R$_e$ we derived from the NICMOS images agree well with those derived by 
Van Dokkum et al. (2003) from observations in the F814W filter.
Our estimate of R$_e$ for the galaxy HUDF\_996
agrees with the one of Daddi et al. (2005) made on the F850W image 
while it is a factor two the estimate of Cimatti et al. (2008).
For the galaxy HUDF\_472 at $z\sim1.9$ our estimate is a factor two
both the estimate of Daddi et al. (2005) and that of Cimatti et al. (2008).
It should be noted that, given the redshift of this latter galaxy, 
the filters F850W and F160W  sample the profile at $\lambda_{rest}\simeq2900$ 
\AA\ and at $\lambda_{rest}\simeq5700$ \AA\ respectively, 
two wavelength ranges which differ substantially for the contribution from 
the young and the old stars respectively.
Thus, in this case the different estimates could reflect a real 
difference of the galaxy profile if observed at such different  
wavelengths (e.g. McGrath et al. 2008).
However, we note that the bi-dimensional fit to the observed images
of the two HUDF galaxies (together with those of the two RDCS galaxies 
with R$_e>8$ Kpc) present significant residuals and thus their derived R$_e$
could be affected by large errors.
The ten SA\# galaxies selected from the GDDS have been recently studied by
Damjanov et al. (2008).
We notice that the effective radii they derive are systematically smaller 
than ours and that for 3 galaxies they derive effective radii as small as 
0.3, 0.4 and 0.7 Kpc respectively, while we never obtain effective
radii smaller than 1 Kpc.
{ In Fig. 2 the comparison between the original estimate of the effective 
radii R$_e^{others}$ [Kpc] obtained by the other groups and our new estimate 
is shown.
The different symbols refer to the different data sets as detailed in the 
caption of the figure.
The original estimate of the effective radii $r_e^{others}$ [arcsec] as
derived by the other groups are also reported in Tab. 2.}

\begin{figure}
\begin{center}
\includegraphics[width=9cm]{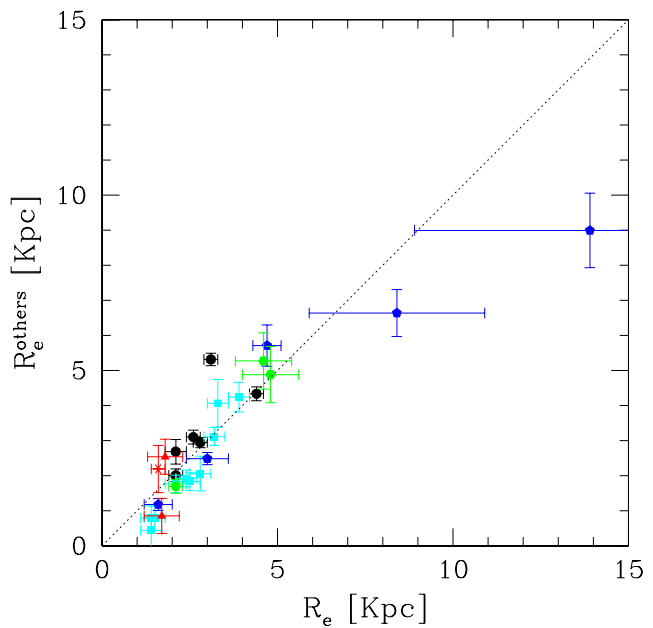}  
\caption{Comparison between the effective radius R$_e$ [Kpc] we obtained from
the analysis of the NICMOS images and the one obtained by the other groups 
 R$_e^{others}$.
Cyan squares are the SA\# ETGs compared with the estimates of Damjanov et al. (2008);
blue pentagons are the CIG\# ETGs in the RDCS 0848+4453 compared with the estimates of
Moriondo et al. (2000); green hexagon are the HDF\# ETGs compared with the estimates
of Stanford et al. (2004); red triangles are the HUDF\# ETGs compared with the
ACS-based estimates of Cimatti et al. (2008); red starred symbol is 53W091 
compared with the estimate of Weddington et al. (2002) and the black points are
6 out of the 10 galaxies studied in Longhetti et al. (2007) compared with
the recent estimates of Damjanov et al. (2008).}. 
\end{center}
\end{figure}

\subsection{Absolute magnitudes, stellar masses and age}
For each galaxy of
the sample  we derived the R-band absolute magnitude M$_R$, the stellar mass 
$\mathcal{M}_*$ and the mean age $Age$ of the stellar population.
We made use of the stellar population synthesis models of 
Charlot and Bruzual (2008, CB08) and of the best-fitting 
code \texttt{hyperz} 
(Bolzonella et al. 2000) to find the best-fitting template
to the  observed spectral energy distribution  (SED) at the 
redshift of each galaxy.
The set of templates considered includes three star formation histories 
(SFHs) described by an exponentially declining star formation rate 
$SFR\propto e^{-t/\tau}$ 
with e-folding time $\tau=[0.1,0.3,0.6]$ Gyr  and two metallicity 
0.4$Z_\odot$ and $Z_\odot$.
We assumed Chabrier initial mass function (IMF; Chabrier et al. 2003).
Extinction A$_V$ has been considered  and treated as a free 
parameter in the fitting.
We adopted the extinction curve of Calzetti et al. (2000) and
we allowed A$_V$ to vary in the range $0\le$A$_V\le0.6$.
For 24 out of the 32 galaxies the best-fitting template is defined by
SFHs with $\tau\le0.3$ Gyr and $A_V\le0.3$.

The R-band absolute magnitude M$_R$ has been derived from the observed flux
in the F160W filter since it samples the R-band in the rest-frame of the 
galaxies.
To derive M$_R$  we used the relation
\begin{equation}
M_R(z)=F160W_{tot} - 5log[D_L(z)]-k_{R,F160W}(z)
\end{equation}
where F160W$_{tot}$ is the total magnitude, $D_L(z)$
is the luminosity distance [Mpc] at the redshift $z$ of the galaxy and 
$k_{R,F160W}(z)$ is the color k-correction term defined as
\begin{equation}
k_{R,F160W}=[R(z=0)- F160W(z)]_{temp}
\end{equation}
where the two magnitudes $R(z=0)$  and $F160W(z)$ are derived from 
the best fitting template at $z=0$ and redshifted 
at the redshift of the galaxy respectively.
The uncertainty affecting this k-correction is typically comparable or smaller
than the photometric errors since the filter F160W is extremely close to the 
rest-frame R-band over the whole redshift range considered and thus the 
dependence on the best fitting template tends to vanish.
{Indeed we have verified that
even considering the oldest template among the best fitting templates 
(the one 4.0 Gyr old) and the youngest template (1 Gyr old)  
the difference between the k-correction  
is less than 0.15 mag over the whole redshift range considered.
Thus, even hypothesizing to fail the fit to
the observed SED of a galaxy we would wrong its absolute magnitude  
by no more than 0.1 mag}.
For this reason we consider our estimate of the R-band absolute magnitude 
of our galaxies extremely reliable. 
      
The stellar mass $\mathcal{M}_*$ of the galaxies we derived is
the one usually computed in the literature and it is given 
by the equation
\begin{equation}
\mathcal{M}_*=2\cdot 10^{-17} \cdot b\cdot 4\pi D^2_L(z)\cdot
\mathcal{M}_*^{model}/L_{\odot}
\end{equation}
where $b$ is the normalization factor of the best fitting model provided
by \texttt{hyperz} and $\mathcal{M}_*^{model}$ is the mass associated
to the best fitting template considering only the stellar mass still
locked into stars.
 This quantity is listed in column 7 of the .4color files of CB08 models.
The mass we derived is the mass locked 
into stars at the epoch of their observation after the gas fraction returned 
to the interstellar medium.
This mass is typically about 60\% the one derived including the gas 
return fraction, i.e. the one obtained by integrating the SFR over 
the age of the galaxy.
A detailed  comparison between different stellar mass estimators
is given by Longhetti et al. (2008) which provide also the 
relations to convert an estimate to another accounting for
different IMFs. 
The uncertainty affecting our stellar mass estimate depends mainly
on the uncertainty affecting the SFH ($\tau$) and the age of the 
best-fitting model and on the best-fitting parameter A$_V$,
three parameters tightly linked among them.
The SFH and the age affect mainly the value of $M_*^{model}$ while
the extinction A$_V$ affect mainly the normalization $b$ of the fit.
Since we deal with galaxies of known spectral type and redshift
the best-fitting SFH and age are sharply constrained producing negligible
differences in the values of $M_*^{model}$.
For instance, even considering the youngest and the oldest ages possible
in this range of redshift, i.e. 1 Gyr and 4 Gyr, 
the corresponding $M_*^{model}$ for the same SFH would differ only by 
a factor 1.15.
On the contrary, the normalization $b$ of the model which is a free
parameter in the fitting, can vary up to a factor $\sim$2 since it depends 
on the photometric accuracy in the various bands, on the number 
of photometric points sampling the spectrum of the galaxy and on the
free parameter A$_V$. 
Thus, the internal accuracy of our stellar mass estimates is within 
a factor two and it is two times the uncertainty affecting the 
absolute magnitudes.
This internal error does not consider the possible systematics
due to different IMFs (e.g. Salpeter IMF provides higher stellar masses 
than Kroupa and Chabrier IMFs) or different library models 
(see e.g. Maraston et al. 2006; Longhetti et al. 2008).  
Thus, for a comparison with other samples such possible systematics
should be taken into account by scaling, if necessary, the different
estimates.
Systematics and scaling relations among different library codes 
can be found in Longhetti et al. (2008). 

The mean age of the best fitting model  
depends both on the SFHs and on the A$_V$ in the way that higher A$_V$ 
and shorter SFHs provide younger best fitting models for a given observed
spectral energy distribution.
As previously said, most of the galaxies (24 out of 32) are best-fitted 
by the shorter SFHs considered, i.e. by models with $\tau\le0.3$ Gyr and 
A$_V\le0.3$ suggesting that the best fitting does not tend toward either
old or young models, 
old with respect to the age of the universe at the redshift
of the galaxy.
However, the main degeneracy is between the age of the best-fitting model
and the extinction A$_V$.
Indeed, while two different SFHs, for instance the one with $\tau\le0.1$ and 
the other $\tau\le0.3$, accounts for differences in the best-fitting age of 
the order of few tenth of Gyr, different values of A$_V$ even within the 
range $0<$A$_V<0.6$ can produce differences as large as 1 Gyr.
An old stellar population can, in fact, be fitted by a young model
reddened by an extinction A$_V>0$.
In order to verify the absence of any systematics in our results
due to the best fitting procedure we have compared the values of the 
extinction A$_V$ with the age of the best fitting template.
In Fig. 3, we plot A$_V$ as a function of the age of 
the best fitting model for the sample of 32 ETGs. 
No systematics are present among the two parameters confirming
that young best fitting templates are not a faked result of
the fitting procedure.
It is worth noting that, with the exception of two galaxies of the GDDS
sample, our estimate of the ages  agrees within
$\sim0.3$ Gyr (0.5 Gyr for the old ones) with the ages derived by the 
various authors from the spectral features of the galaxies.
 
 \begin{figure}
\begin{center}
\includegraphics[width=9cm]{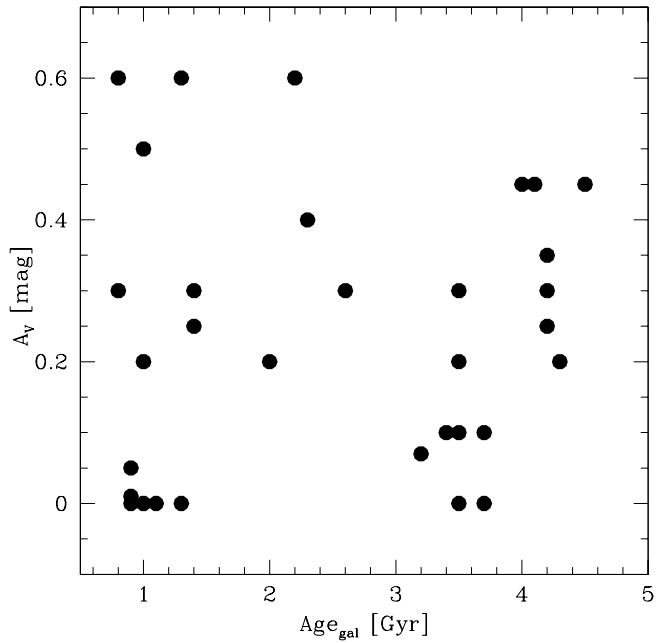}  
\caption{The extinction A$_V$ is plotted as a function of the mean age 
Age$_gal$ of the best fitting model for the sample of 32 ETGs. 
No systematic are present between the two parameters.}
\end{center}
\end{figure}


In Tab. 2 we report for each galaxy the R-band absolute magnitude, 
the stellar mass and the mean age derived for each galaxy
from the best fitting to the photometry.

\section{The evolution of the Kormendy relation}
The Kormendy relation (KR, Kormendy 1977) is a linear scaling relation 
between the logarithm of the effective radius R$_e$ [Kpc], i.e. 
the radius containing half of the light, and the mean 
surface brightness $\langle\mu\rangle_e$  [mag/arcsec$^2$]:
\begin{equation}
\langle\mu\rangle_e = \alpha + \beta \log(R_{e})
\end{equation}

The ETGs follow  this relation with 
a fixed slope $\beta\sim3$ up to $z\sim1$ (e.g. Di Serego et al. 2005)
while the zero point $\alpha$ varies
with the redshift  reflecting the evolution that the galaxy underwent.
\begin{figure*}
\begin{center}
\includegraphics[width=8.8cm]{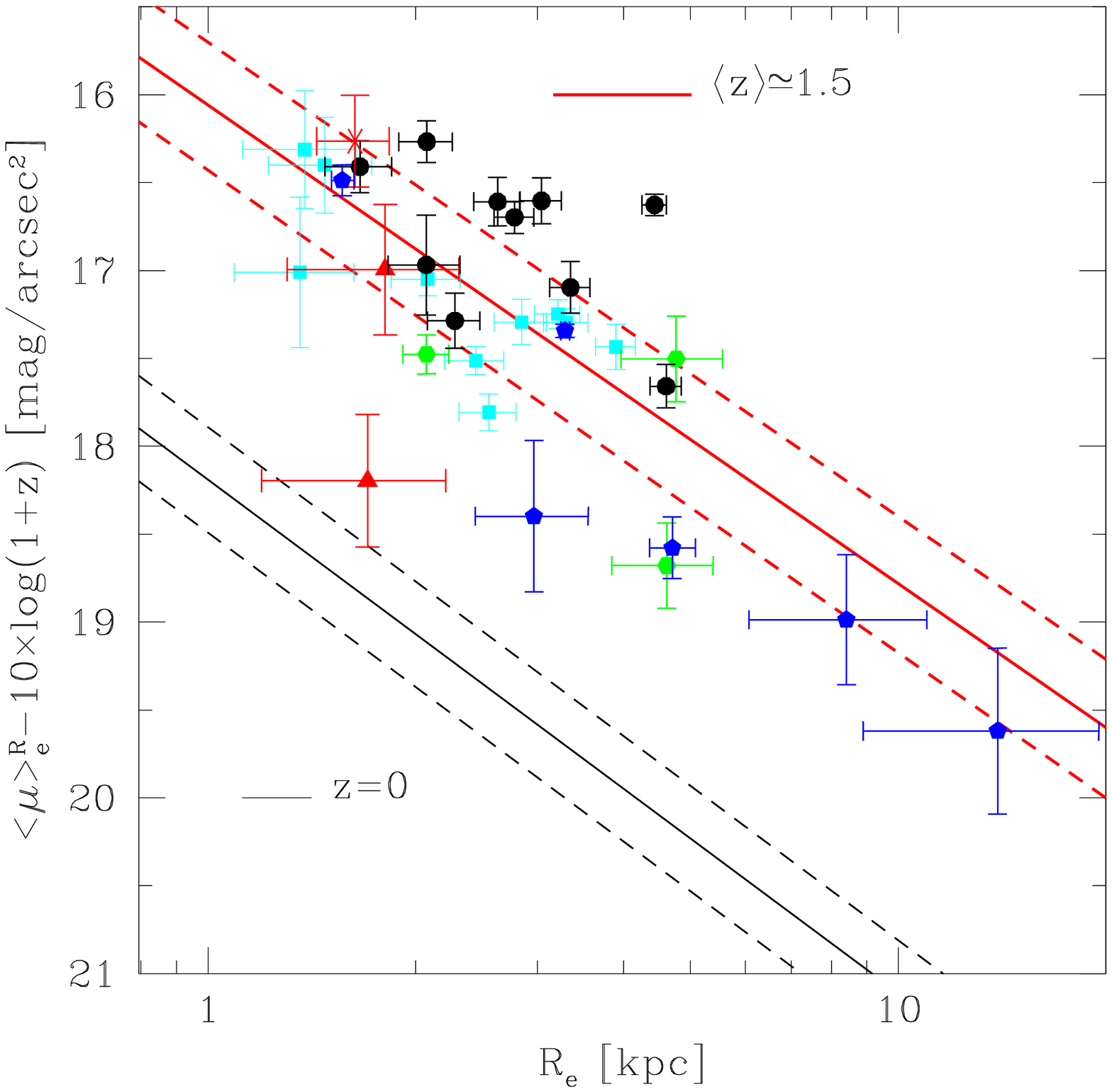}  
\includegraphics[width=8.8cm]{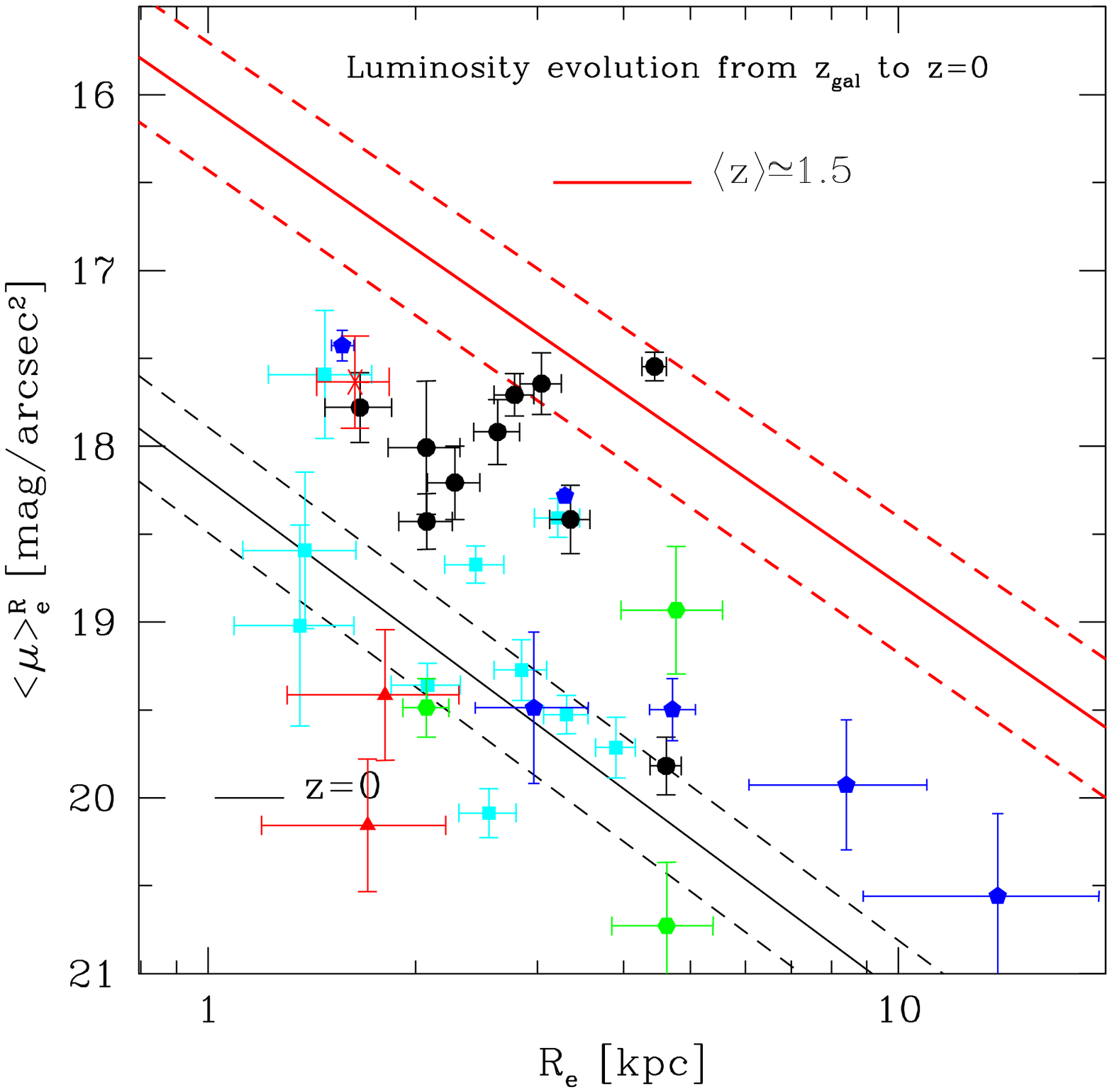}
\vskip -0.5truecm
\caption{{\em Left panel} - Mean surface brightness 
$\langle\mu\rangle_e$ versus  effective radius R$_e$ [kpc]
for the 32 ETGs of the sample.
The thin (black) solid line represents the KR at $z\sim0$ (eq. 7) and
the short-dashed lines represent  the $\pm1\sigma$ dispersion of the relation.
The thick (red) line is the KR at $z\sim1.5$ as resulting from the best fit
to the sample (eq. 8). 
The short dashed lines represent the scatter around this relation. 
All the data have been corrected for the cosmological dimming factor 
$(1+z)^4$ thus, the deviation from the KR at $z=0$ reflects 
the evolution of the SB due to the luminosity and/or size evolution 
of galaxies. 
Black points are the S2\# ETGs (Longhetti et al. 2007), cyan squares are
the SA\# ETGs from the GDDS (McCarthy et al. 2004), blue pentagons are the
CIG\# ETGs in the RDCS 0848+4453 (Stanford et al. 1997), green hexagon are the 
HDF\# ETGs, red triangles are the HUDF\# ETGs (Cimatti et al. 2008) and the red
starred symbol is 53W091 (Dunlop et al. 1996). 
{\em Right panel} - Mean surface brightness 
$\langle\mu\rangle_e$ versus  effective radius R$_e$ [kpc] for the 32
galaxies of our sample after than they have evolved in luminosity. 
Each galaxy has evolved in luminosity over the interval
$\Delta t$ corresponding to the time elapsed from the redshift $z$ 
of the galaxy to $z=0$.
The evolution $E(z)$ has been derived accounting for the different age of 
each galaxy at the observed redshift (see Sec. 4).
}
\end{center}
\end{figure*}

In Fig. 4 (left panel) the values of R$_e$ and $\langle\mu\rangle_e^R$ in 
the R band for
the 32 galaxies at $1<z<2$ derived as described in the previous section
are plotted on the [$\mu_{e}$,R$_{e}$] plane.
The values of $\langle\mu\rangle_e^R$ have been corrected for the 
cosmological dimming factor $(1+z)^4$.
We converted the surface brightness $\langle\mu\rangle_e^{F160W}$
in the F160W filter into that  in the 
rest-frame R-band $\langle\mu\rangle_e^{R}$ applying to each galaxy the 
k-correction $k_{R,F160W}$ described in the Sec. 3.2.
In Fig. 4, the observed KR in the R band at $z\sim0$ 
(see La Barbera et al. 2003)
\begin{equation}
\langle\mu\rangle_e=18.2+2.92 log(R_e),\hskip 2truecm z=0
\end{equation}
is also shown (thin black line).
The thick (red) solid line is the KR we obtain from the best fitting of our 
sample
\begin{equation}
\langle\mu\rangle_e=16.1^{+0.1}_{-0.2}+2.72^{+0.5}_{-0.2}log(R_e),
\hskip 1truecm 1<z<2.
\end{equation} 
and the two thick dashed lines represent the dispersion at 1$\sigma$ of the
relation.
It can be seen that the slope $\beta$ of the KR we
fit at  $z\sim1.5$ does not change significantly  with respect to the KR
at $z=0$ while we detect a significant evolution of the zero-point 
$\alpha$ which changes more than 2 magnitudes in this redshift range.
Thus, at least up to $z\sim1.5-2$ ETGs tend to distribute on the
[$\mu_{e}$,R$_{e}$] plane according to a KR with a slope similar to
the one of local ETGs $z=0$.
The different zero-point accounts for the evolution which undergo the galaxies
and tell us that in case of pure luminosity evolution, i.e. at
constant R$_e$, galaxies must evolve by more than 2 magnitudes in
the rest-frame R band from $z_{mean}\sim1.5$ to $z=0$,
in agreement with previous results (see e.g. McIntosh et al. 2005;
Longhetti et al. 2007; Cimatti et al. 2008). 
Such evolution exceeds almost 1 mag the one expected assuming an average 
passive luminosity evolution (PLE) for the whole sample
and exceeds much more the evolution inferred from the 
 observed luminosity function of galaxies in this
redshift range (e.g. Feulner et al. 2007; Cirasuolo et al. 2007; Marchesini et al. 2007; 
Zucca et al. 2006; Pozzetti et al. 2003).
However, our sample of galaxies is spread over a range of redshift, $1<z<2$, 
which spans about 2.5 Gyr of time, a large interval if compared
to the cosmic time at $z\sim1$ (5.7 Gyr) and at $z\sim2$ (3.2 Gyr). 
Thus, in order to assess whether the observed evolution of the zero point
$\alpha$ of the KR can be accounted for by luminosity evolution, we derived
for each galaxy its own luminosity evolution $E(z)$ in the rest-frame R-band  
over the interval $\Delta t=t(z=0)-t(z)$,
corresponding to the time elapsed from $z$ to $z=0$.
The evolutionary term $E(z)$ has been calculated taking into account the 
different ages of the galaxies at the observed redshift as provided by 
the best-fitting model.
Thus, for each galaxy we computed the term
$E(z)=[M_R(z)-M_R(z=0)]_{model}$
i.e. the difference between the R-band absolute magnitude of the best 
fitting model with age $Age$ and the R-band magnitude of the same model 
with age $Age+\Delta t$.
This term added to equation (3) provides the surface brightness
$\langle\mu\rangle_e^R$ that our galaxies would have at $z=0$ in the 
case of pure luminosity evolution. 
In Fig. 4 (right panel) we show how the 32 ETGs of our sample would be 
displaced at $z=0$ in the [$\mu_{e}$,R$_{e}$] plane.
{ We see that almost a half of the sample reaches a surface brightness
not exceeding the one derived from the KR at $z=0$ while the remaining half 
shows a surface brightness in excess of $\sim$1 magnitude with respect to 
the local KR.
In particular, the luminosity evolution brings  11 ETGs onto the KR at $z=0$
and two ETGs just below it while leaves the remaining 19 ETGs with a 
surface brightness exceeding much more than one sigma the local KR.}
It seems that for a fraction of ETGs at $z>1$ the expected luminosity
evolution is sufficient to account for their surface brightness.
They move in the [$\mu_{e}$,R$_{e}$] plane from high to low
redshift in agreement with the local KR.
On the contrary, for the remaining fraction of ETGs the expected luminosity
evolution is not sufficient to dim their surface brightness 
to the one defined by the local KR.
The other parameter involved in the KR relation, 
the effective radius R$_e$, 
must evolve and the hypothesis of fixed size must be rejected
for these galaxies.
A size evolution of at least a factor $\sim$2.5 from $z\sim1.5$ to $z=0$
 is needed to account for the observed surface brightness excess.

\section{Young {\it VS} old ETGs: two distinct populations at $z\sim1.5$ ?}
In the previous section we have seen that for 13 ETGs the pure luminosity
evolution can move them from the KR at $z\simeq1.5$ onto the KR at $z=0$,
while for the remaining 19 ETGs a different more complex evolution is 
required.
In fact, this suggests that two distinct populations of ETGs exist 
at $z\simeq1.5$. 
In order to better investigate this evidence, we compared the properties
of the 13 ETGs whose luminosity evolution places them onto
the KR at $z=0$ with the remaining 19 ETGs. 
Basically, the luminosity evolution $E(z)$ derived by models over a given 
interval $\Delta t$ of time depends on the SFH  and on the age of the 
best-fitting model. 
The SFHs which fit the SED of our galaxies are described by e-folding time 
$\tau$ much shorter than the typical $\Delta t$ over which the luminosity 
evolution is computed ($\tau\le0.3$ Gyr to be compared with 
$\Delta t\sim9$ Gyr from $z=1.5$ to $z=0$).
Thus, the slightly different values of $\tau$ cannot produce significant 
differences in $E(z)$ over this interval.
Consequently, the other parameter affecting the luminosity 
evolution, that is the age of the best fitting model, must be the reason of
the different behaviour of the two sub-samples.
If so, we expect that the two sub-samples of galaxies show a
different age distribution.

In Fig. 5  the age distribution of the 13 galaxies 
(dashed green histogram) which agree with the KR at $z=0$ is compared
with the age distribution of the remaining 19 ETGs (solid red histogram).
It is worth noting that the two sub-samples, with the exception of 2 ETGs 
(one for each sub-sample), describe two separated distributions. 
The first distribution is sharply picked at $\sim1$ Gyr, the other  
distribution picks at $\sim3.5$Gyr.
As shown in Sec. 3.2 age and extinction A$_V$,
are not correlated and no systematic are present between
them (see Fig. 3).
Moreover, the different values of $\tau$ considered cannot account
for such different ages.
Indeed, a galaxy 1 Gyr old fitted by a SFH $\tau=0.1$ Gyr would be
$\sim1.3$ Gyr old if fitted with a model $\tau=0.3$ Gyr.
Analogously, a galaxy 4 Gyr old fitted by a SFH $\tau=0.3$ Gyr would be
$\sim3.5$ Gyr old if fitted with a model $\tau=0.1$ Gyr.
Thus, the two different distributions are not
a consequence of the degeneracy between SFH, age and 
extinction but they reflect  real differences among
the ETGs:
the 13 ETGs which fall on the local KR are, in fact, 
younger than the remaining 19 ETGs.
Given the uncertainties discussed above, it is reliable to consider
a mean difference of about 1.5-2 Gyr between the age of the two populations.
Hereafter, we will  refer to these two populations as young ETGs (yETGs)
and old ETGs (oETGs).
\begin{figure}
\begin{center}
\includegraphics[width=9cm]{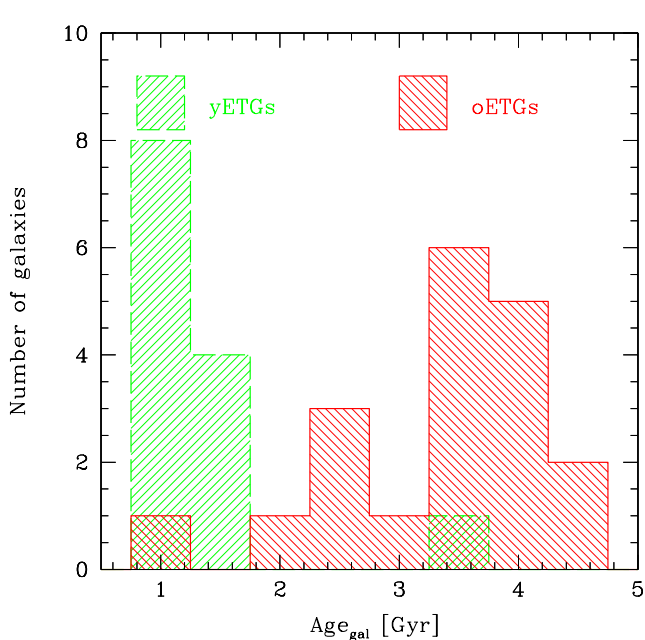} 
\vskip -0.5truecm
\caption{Age distributions of the two sub-sample of ETGs. 
The dashed (green) histogram represents the distribution of the 13 ETGs
whose surface brightness agrees with the local KR in case of luminosity
evolution. The solid (red) histogram represents the distribution of
the remaining 19 ETGs whose SB exceeds by more than one $\sigma$ the KR.
}
\end{center}
\end{figure}

\begin{figure}
\begin{center}
\includegraphics[width=9cm]{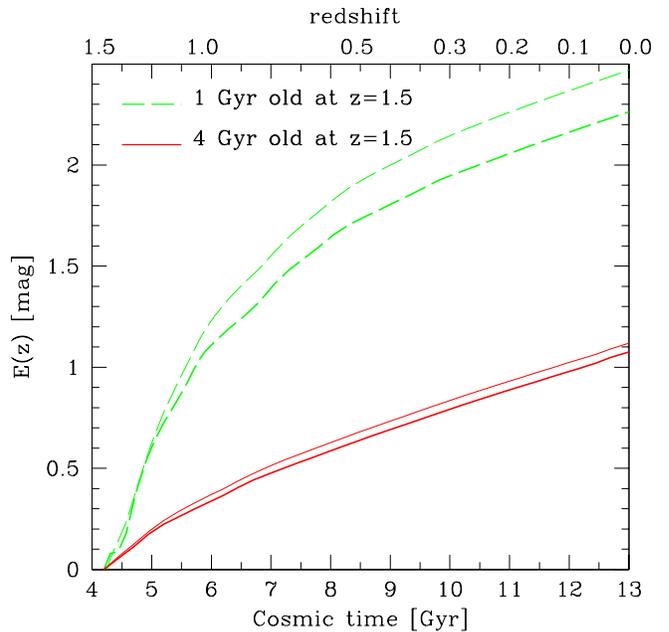} 
\vskip -0.5truecm
\caption{
Evolution term $E(z)$ as a function of the cosmic time (bottom x-axis)
or redshift (upper x-axis) for two galaxies which at $z=1.5$ are
1 Gyr old (dashed green line) and 4 Gyr old (solid red lines) respectively.
The thin lines have been obtained with a SFH described by $\tau=0.1$ Gyr, 
the thick lines refer to a model with $\tau=0.3$ Gyr.  
}
\end{center}
\end{figure}

In fact, the different age of the stellar populations of the two sub-samples
of ETGs is the reason of their different behaviour with respect
to the KR. 
Figure 6 displays the evolution term $E(z)$ in the R-band
as a function of the cosmic time (bottom x-axis) or redshift (upper x-axis) 
for two galaxies whose stellar populations at $z=1.5$ are 1 Gyr old 
(dashed red line) and 4 Gyr old (solid red line).
The thin lines refer to a SFH with $\tau=0.1$ Gyr while the thick
lines refer to a $\tau=0.3$ Gyr.
In fact, we see that the difference in the luminosity evolution $E(z)$ between 
the old and the young stellar population is about 1 mag at $z=0$ 
independently of the different values of $\tau$.
Thus, the different mean age of the stellar populations of ETGs at $z\sim1.5$
is the reason of their different expected evolution.
We have compared also the absolute magnitude distribution and the stellar 
mass distribution of these two populations to gather other information 
about their evolutionary status and  to search for signs of different history
of star formation and mass assembly.
In Fig. 7 we show the distributions of the absolute magnitude (left panel)
and of the stellar mass (right panel) of the yETGs (dashed red histogram) and
of the oETGs (solid red histogram).
We see that yETGs tend to be less luminous and, accordingly, less massive
than the oETGs even if the effect is not statistically significant,
as confirmed by the KS-test performed to compare the
two distributions ($P(D>D_{max})=0.02$). 
However, it can be seen that the high-luminosity/mass
tail is populated only by old ETGs while the low-luminosity/mass
tail is composed of only young ETGs.

The SED fitting of our ETGs is based on optical and near-IR photometry
which at $z\sim1.5$ samples the wavelength range $\lambda<0.8$ $\mu$m.
In practice, at this redshift, 6 out of the 8-9 photometric points 
sample the UV and blue rest-frame emission of the galaxies whose
continuum shape is affected by star formation episodes
even if involving a negligible fraction of the stellar mass (see e.g. 
Fig. 2 in Longhetti et al. 2008).
For this reason, it is more appropriated to consider the age we derived 
as a lower
limit to the time elapsed since the last episode of star formation.
If this latter is the major one, then this age will provide
the formation redshift of the stellar population.
Given the different  age of the two populations of ETGs 
it is likely that the formation redshift of their stellar populations
are different.
In particular, the stellar component of the oETGs ($Age_{med}\sim3.5$) formed 
at $z_f\sim5-6$ while the stellar populations
of yETGs ($Age_{med}\sim1$) formed at $z_f>2.5-3$, or at least the youngest
population. 

\begin{figure*}
\begin{center}
\includegraphics[width=8.8cm]{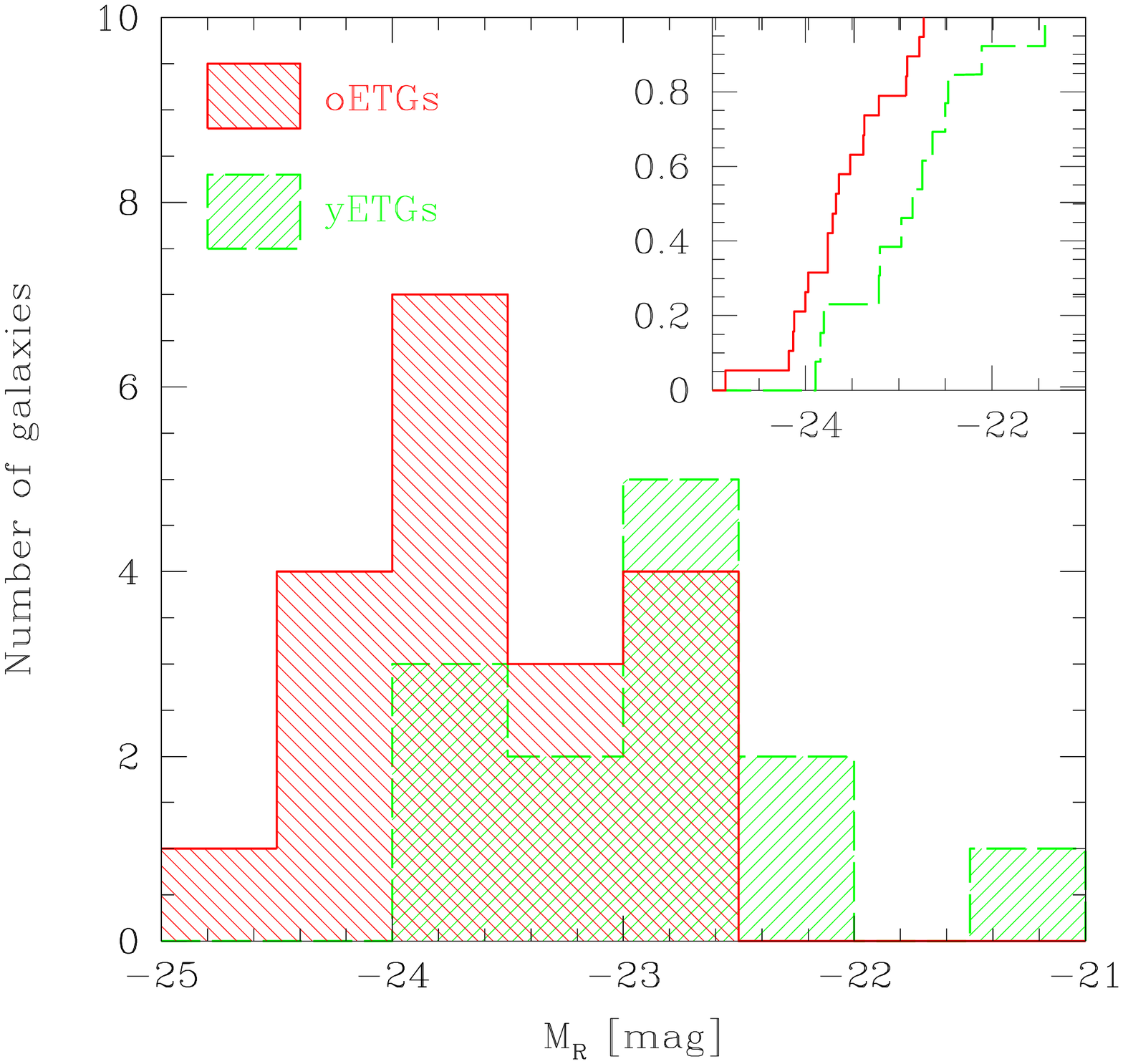} 
\includegraphics[width=8.8cm]{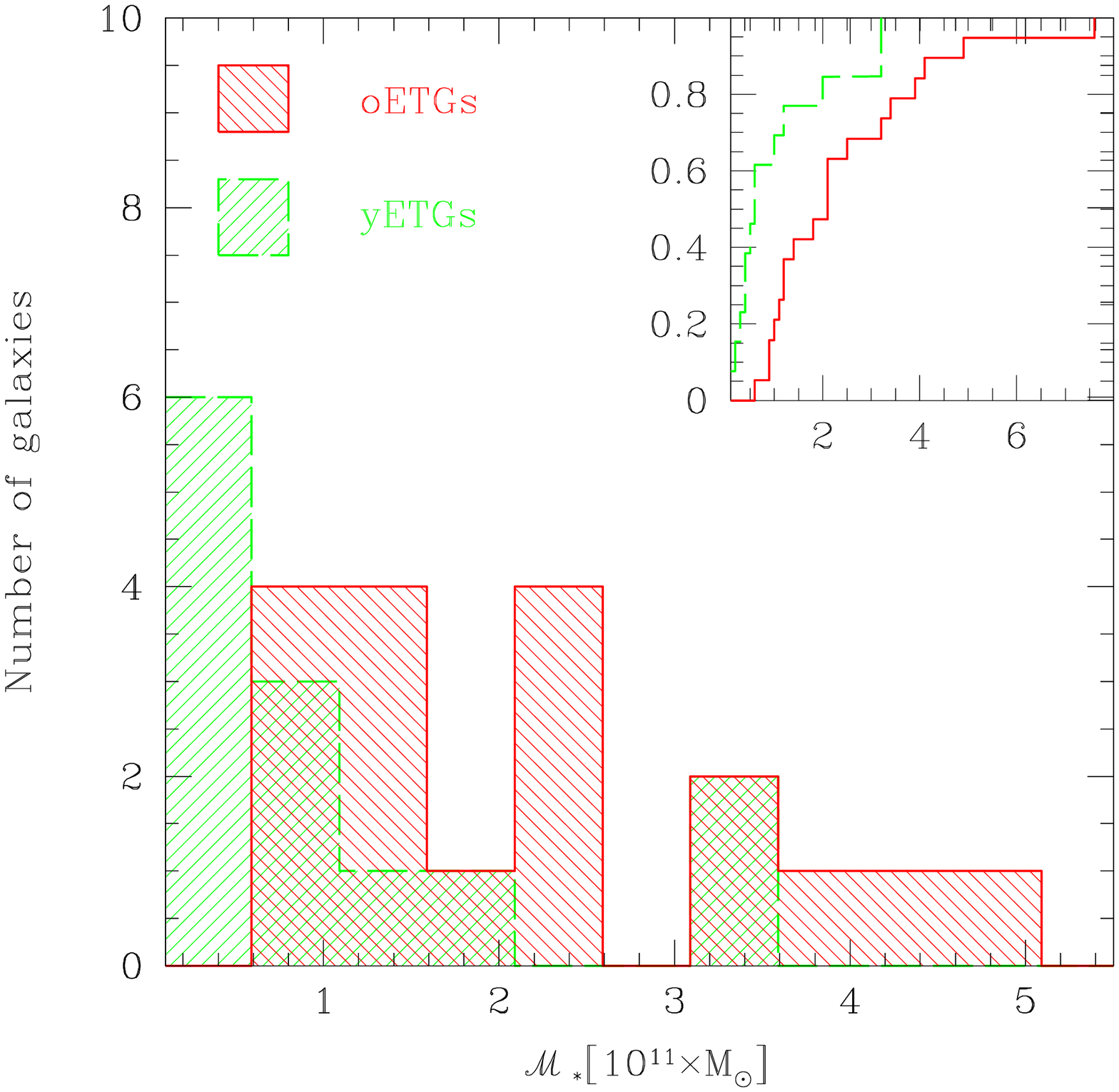} 
\vskip -0.5truecm
\caption{Distribution of the R-band absolute magnitude M$_R$ (left panel)
and of the stellar mass $\mathcal{M}_*$ for the 13 yETGs (dashed green histogram) 
and for the 19 oETGs  (solid red histogram). The inside panels show cumulative
distributions. The KS test performed provides a probability that the two
distributions come from the same population $P\simeq0.02$.
}
\end{center}
\end{figure*}

\begin{table*}
\caption{Morphological parameters of galaxies. The effective radii obtained
by fitting the Sersic profile and the resulting surface brightnesses take
into account the correction for the underestimate of $r_e$ derived from the 
simulations (see Sec.3.1). 
We applied a correction of 0.07 arcsec to the NIC2 data and of 0.03 arcsec 
to the NIC3 data.
The error on the effective radii takes into account both the formal error
of the profile fitting and the rms observed in recovering the intrinsic radius
of simulated galaxies, namely $\sigma^{NIC2}_{r_e}=0.02 $ arcsec and 
$\sigma^{NIC3}_{r_e}=0.04 $ arcsec.  
The magnitude F160W$_{tot}$ is the total magnitude of the best fitting profile
as derived by \texttt{galfit}.
The effective radius r$_e^{others}$ is the original effective radius estimated by
other groups.
}
\centerline{
\begin{tabular}{lcccccccccc}
\hline
\hline
  Object  & $z$ & F160W$_{tot}$& M$_{R}$ & r$_e$ & r$_e^{others}$& R$_e$ & $\langle\mu\rangle_e^R$
  &$\langle\mu\rangle_e^{F160W}$ & Age & $\mathcal{M}_*$ \\
          &    &[mag]      &[mag]   & [arcsec] & [arcsec] & [Mpc] & [mag/arcsec$^2$]&
	  [mag/arcsec$^2$] & [Gyr] & [10$^{11}$ M$_{\odot}$] \\
  \hline
S2F5\_109 & 1.22  &17.47$\pm$0.02  & -24.86 &  0.53$\pm$0.02   & --  &  4.4$\pm$0.2    &  20.1$\pm$0.1   &  18.1$\pm$0.1  &  3.5   & 7.6  \\
S7F5\_254 & 1.22  &19.46$\pm$0.03  & -22.91 &  0.27$\pm$0.02   & --  &  2.3$\pm$0.2    &  20.8$\pm$0.2   &  18.6$\pm$0.2  &  4.5   & 3.9  \\
S2F1\_357 & 1.34  &18.72$\pm$0.03  & -23.71 &  0.33$\pm$0.02   & --  &  2.8$\pm$0.2    &  20.4$\pm$0.2   &  18.3$\pm$0.2  &  4.2   & 4.9  \\
S2F1\_389 & 1.40  &19.79$\pm$0.03  & -22.73 &  0.25$\pm$0.03   & --  &  2.1$\pm$0.3    &  20.8$\pm$0.2   &  18.7$\pm$0.2  &  3.5   & 1.8  \\
S2F1\_511 & 1.40  &19.15$\pm$0.03  & -23.64 &  0.25$\pm$0.02   & --  &  2.1$\pm$0.2    &  20.1$\pm$0.2   &  18.1$\pm$0.2  &  1.0   & 0.9  \\
S2F1\_142 & 1.43  &18.65$\pm$0.03  & -24.00 &  0.36$\pm$0.02   & --  &  3.1$\pm$0.2    &  20.5$\pm$0.2   &  18.4$\pm$0.2  &  3.5   & 4.1  \\
S7F5\_45\ & 1.45  &18.83$\pm$0.03  & -23.89 &  0.55$\pm$0.03   & --  &  4.7$\pm$0.3    &  21.5$\pm$0.1   &  19.5$\pm$0.1  &  1.0   & 2.0  \\
S2F1\_633 & 1.45  &19.00$\pm$0.03  & -23.67 &  0.31$\pm$0.02   & --  &  2.6$\pm$0.2    &  20.5$\pm$0.2   &  18.5$\pm$0.2  &  2.6   & 2.5  \\
S2F1\_443 & 1.70  &19.44$\pm$0.03  & -23.76 &  0.40$\pm$0.03   & --  &  3.4$\pm$0.3    &  21.4$\pm$0.2   &  19.4$\pm$0.2  &  3.2   & 3.4  \\
S2F1\_527 & 1.35  &19.50$\pm$0.03  & -23.38 &  0.20$\pm$0.03   & --  &  1.7$\pm$0.3    &  20.2$\pm$0.2   &  18.0$\pm$0.2  &  2.3   & 1.1  \\
SA12-5592& 1.623  &20.30$\pm$0.01  & -22.85  & 0.16$\pm$0.04   &0.05$\pm$0.05 &   1.4$\pm$0.3   &	20.50$\pm$0.4 & 18.36$\pm$0.4  &  0.9  & 0.3  \\
SA12-5869& 1.510  &19.53$\pm$0.02  & -23.20  & 0.34$\pm$0.04   &0.25$\pm$0.06 &   2.8$\pm$0.3   &	21.29$\pm$0.3 & 19.16$\pm$0.3  &  1.2  & 0.4  \\
SA12-6072& 1.576  &20.96$\pm$0.01  & -22.11  & 0.16$\pm$0.04   &0.09$\pm$0.04 &   1.4$\pm$0.3   &	21.12$\pm$0.4 & 18.98$\pm$0.4  &  1.4  & 0.3  \\
SA12-8025& 1.397  &19.87$\pm$0.01  & -22.92  & 0.30$\pm$0.04   &0.24$\pm$0.03 &   2.4$\pm$0.3   &	21.31$\pm$0.3 & 19.18$\pm$0.3  &  3.7  & 0.5  \\
SA12-8895& 1.646  &19.20$\pm$0.02  & -23.84  & 0.46$\pm$0.04   &0.50$\pm$0.05 &   3.9$\pm$0.3   &	21.66$\pm$0.2 & 19.56$\pm$0.2  &  0.8  & 0.7  \\
SA15-4367& 1.725  &20.61$\pm$0.01  & -22.64  & 0.30$\pm$0.04   &0.22$\pm$0.03 &   2.5$\pm$0.3   &	22.16$\pm$0.3 & 20.05$\pm$0.3  &  0.9  & 0.4  \\
SA15-5005& 1.845  &20.46$\pm$0.01  & -22.97  & 0.25$\pm$0.04   &0.21$\pm$0.03 &   2.1$\pm$0.3   &	21.59$\pm$0.3 & 19.42$\pm$0.3  &  0.9  & 0.4  \\
SA15-7543& 1.801  &19.64$\pm$0.01  & -23.80  & 0.39$\pm$0.04   &0.48$\pm$0.08 &   3.3$\pm$0.3   &	21.77$\pm$0.2 & 19.60$\pm$0.2  &  1.0  & 0.9  \\
SA22-0189& 1.490  &19.19$\pm$0.01  & -23.76  & 0.38$\pm$0.04   &0.37$\pm$0.03 &   3.2$\pm$0.3   &	21.21$\pm$0.2 & 19.08$\pm$0.2  &  3.5  & 1.8  \\
SA22-1983& 1.488  &20.03$\pm$0.02  & -22.78  & 0.17$\pm$0.04   &0.09$\pm$0.04 &   1.5$\pm$0.3   &	20.36$\pm$0.4 & 18.23$\pm$0.4  &  3.7  & 1.0  \\
CIG\_237 & 1.271  &20.14$\pm$0.03  & -22.50  &  0.35$\pm$0.07  &0.29$\pm$0.02   &   3.0$\pm$0.6  &  21.96$\pm$0.4  &  19.89$\pm$0.4  & 3.5 &  0.3 \\
CIG\_65  & 1.263  &18.85$\pm$0.01  & -24.18  &  0.39$\pm$0.04  & --             &   3.3$\pm$0.3  &  20.89$\pm$0.2  &  18.82$\pm$0.2  & 4.2 &  2.1  \\
CIG\_142 & 1.277  &19.63$\pm$0.02  & -23.21  &  0.19$\pm$0.05  &0.14$\pm$0.02   &   1.6$\pm$0.4  &  20.06$\pm$0.6  &  17.99$\pm$0.6  & 4.2 &  1.0  \\
CIG\_70  & 1.275  &18.02$\pm$0.01  & -24.12  &  1.70$\pm$0.60  &1.10$\pm$0.13   &  13.9$\pm$5.0  &  23.19$\pm$0.8  &  21.12$\pm$0.8  & 4.2 &  2.1  \\
CIG\_108 & 1.277  &18.48$\pm$0.02  & -23.97  &  1.00$\pm$0.30  &0.79$\pm$0.08   &   8.4$\pm$2.5  &  22.56$\pm$0.6  &  20.49$\pm$0.6  & 4.2 &  1.4  \\
CIG\_135 & 1.276  &19.33$\pm$0.02  & -23.52  &  0.56$\pm$0.05  &0.68$\pm$0.07   &   4.7$\pm$0.4  &  22.15$\pm$0.2  &  20.08$\pm$0.2  & 4.3 &  0.9  \\
HDF\_1031 & 1.015 &19.37$\pm$0.03  & -22.47  & 0.26$\pm$0.02   &0.21   &  2.1$\pm$0.2   &  20.52$\pm$0.2  &  18.42$\pm$0.2 &  1.1 &   0.2 \\
HDF\_1523 & 1.050 &17.67$\pm$0.02  & -24.13  & 0.59$\pm$0.10   &0.60   &  4.8$\pm$0.8   &  20.62$\pm$0.4  &  18.52$\pm$0.4 &  2.0 &   2.1 \\
HDF\_ 731 & 1.755 &20.20$\pm$0.05  & -23.21  & 0.55$\pm$0.09   &0.63   &  4.6$\pm$0.8   &  23.08$\pm$0.4  &  20.92$\pm$0.4 &  1.4 &   0.4\\
HUDF\_472 & 1.921 &20.99$\pm$0.03  & -22.75  & 0.20$\pm$0.06   &$^a$0.10(0.08)   & 1.7$\pm$0.5    & 21.65$\pm$0.7   & 19.48$\pm$0.7  &  0.8 & 0.4 \\
HUDF\_996 & 1.390 &21.43$\pm$0.05  & -21.43  & 0.22$\pm$0.06   &$^a$0.31(0.10)   & 1.8$\pm$0.5    & 22.24$\pm$0.6   & 20.13$\pm$0.6  &  1.3 & 0.1 \\
53W091    & 1.55  &19.77$\pm$0.04  & -23.37  & 0.19$\pm$0.02   &0.30$\pm$0.08   & 1.6$\pm$0.2    & 20.33$\pm$0.3   & 18.19$\pm$0.3  &  2.4 & 0.6 \\
\hline
\hline
\end{tabular}
}
$^a$ These values have been derived from HST-ACS observations in the F850W filter.
The values out of the brackets have been derived from Daddi et al. (2005)
those within the brackets from Cimatti et al. (2008). 
\end{table*}

\section{The size-luminosity and the size-mass relations}
The dependence of the characteristic size of galaxies on their luminosity
and on their stellar mass has been recently studied for large samples 
of local galaxies (e.g. Shen et al. 2003), at intermediate redshift 
($z\sim1$, McIntosh et al. 2005) and  at high-redshift ($z<3$) 
 (e.g. Trujillo et al. 2004, 2006b; Cimatti et al. 2008;
  van Dokkum et al. 2008).
 Shen et al. (2003) on the basis of the SDSS data, 
show that the size of  local early types and late types 
increases, as expected, according to their luminosity and stellar mass.
However, the ETGs follow steeper relations than late types showing
that the size of a stellar system is not simply a function of its mass
and that the history of its mass assembly can affect these relations
(Shen et al.2003).
McIntosh et al. (2005) study the evolution of these relations up to 
$z<0.8-1$ by combining GEMS data (Galaxy Evolution from Morphology and SEDs, 
Rix et al. 2004) with COMBO-17 data (Classifying Objects by Medium-Band
Observations, Wolf et al. 2003, 2006b) while Trujillo et al.  (2007)
extend the study up to $z\sim2$ thanks to the DEEP-2 survey (Davis et al. 2003;
Bundy et al. 2006) and to $z\sim3$ (Trujillo et al. 2004, 2005) 
thanks to the FIRES data (Franx et al. 2000).
Our data allow us to define these relations  for a sample
of ETGs at $1<z<2$ with secure spectroscopic redshift and classification
and on the basis of the morphology derived by their red rest-frame continuum.

\begin{figure*}
\begin{center}
\hskip -0.2truecm
\includegraphics[width=8.9cm]{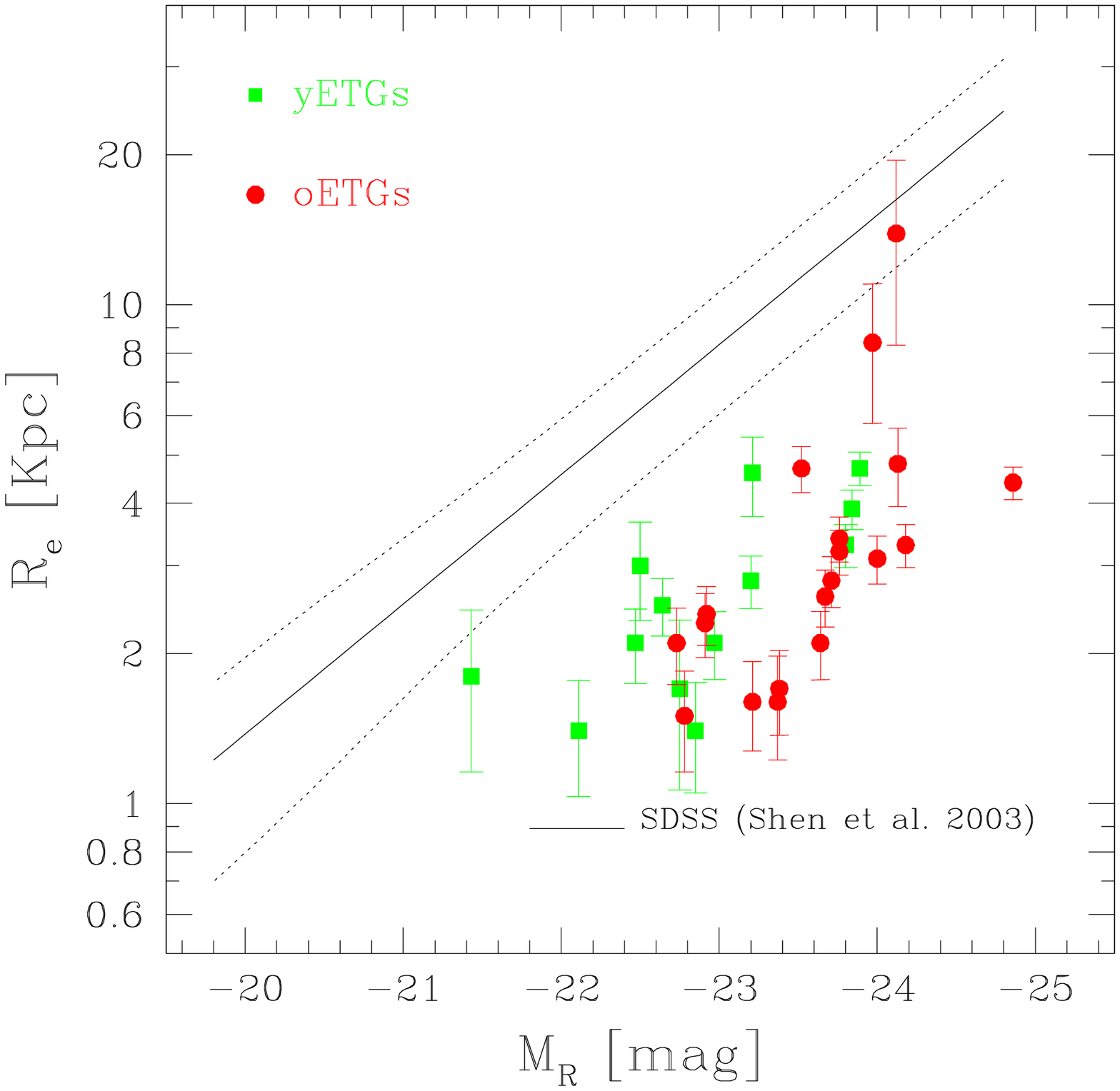} 
\includegraphics[width=8.9cm]{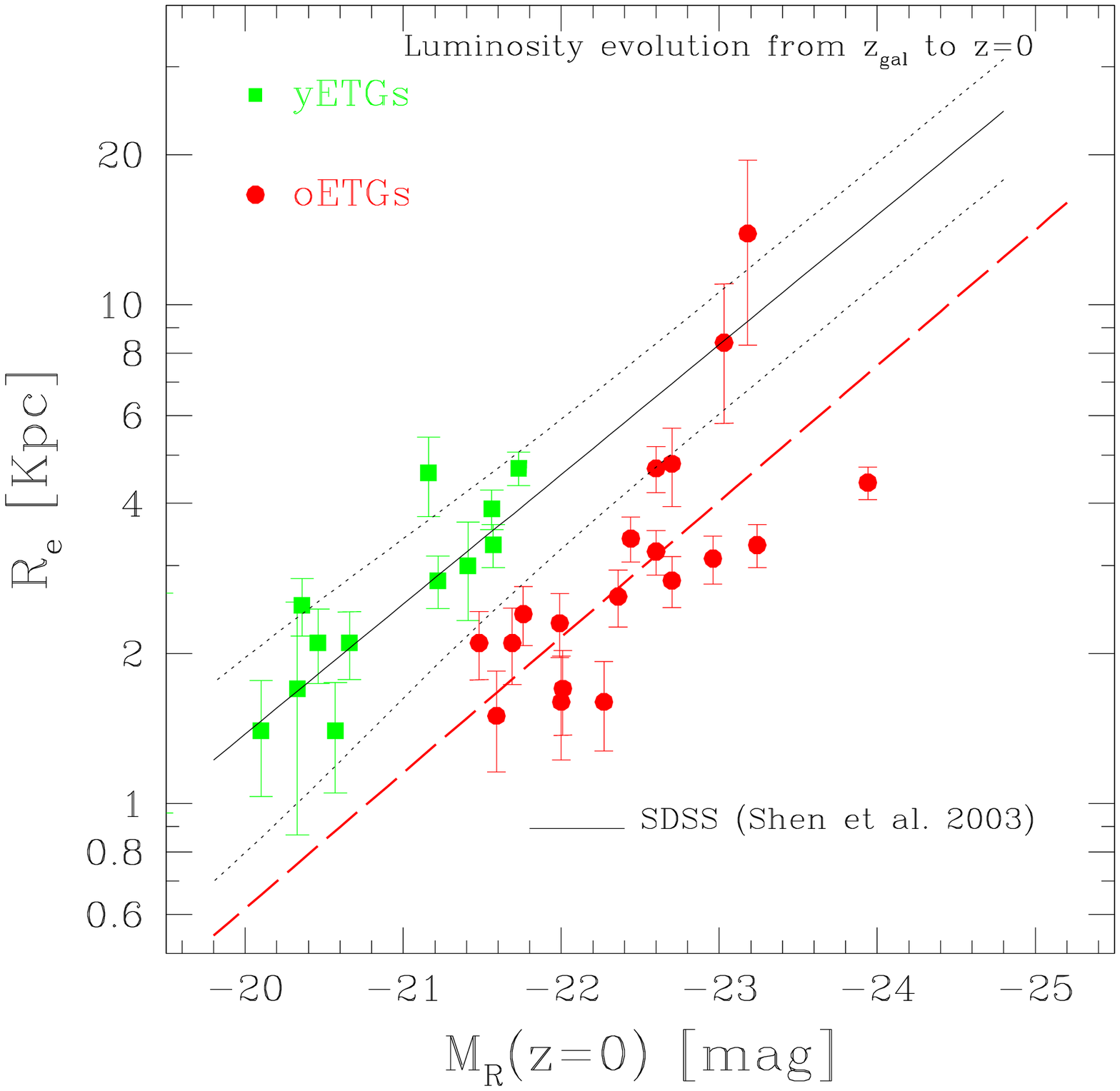} 
\vskip -0.5truecm
\caption{Left panel - Size-luminosity relation for our sample of yETGs 
(green squares)  and oETGs (red points) at $z\sim1.5$ compared with 
the local relation found by Shen et al. (2003, solid line). 
The dotted lines represent the scatter of the relation.
Right panel - Size-luminosity relation in the case of pure luminosity evolution.
It is shown how the 32 ETGs of our sample would be 
displaced at $z=0$ in the [R$_{e}$,$M_R$] plane in case of pure luminosity 
evolution, i.e. the R-band absolute magnitude plotted is 
 $M_R(z=0)=M_R(z)-E(z)$. The dashed (red) line is the best fitting relation
 $logR_e=-0.28M_R(z=0)-5.6$ to the oETGs (see Sec. 6).
}
\end{center}
\end{figure*}

The size-luminosity (S-L) relation for our sample at $z\sim1.5$ is shown in Fig. 8
(left panel) and it is compared with the relation  found in 
the $r$ band  by Shen et al. (2003) using the Sersic half light radius 
for  local ETGs (solid line)
\begin{equation}
log (R_e)=-0.26M_R-5.06
\end{equation}

The dotted lines represent the scatter of the relation.
Young ETGs are marked with (green) squares while old ETGs with (red)
points.
The R-band absolute magnitude M$_R$ of our galaxies is the one at the 
redshift of the galaxies.
The offset with respect to the relation of Shen et al. reflects the evolution 
which ETGs undergo from their redshift to $z=0$, the same evolution
observed in the comparison of the KR shown in Fig. 4 (the one at $z\sim1.5$ 
and that at $z=0$).
It is worth noting that the difference between the $r$ photometric band
of the SDSS and the $R$ Cousins band we use is about 0.2 mag 
(Fukugita et al. 1995).
Thus the possible uncertainties related to the transformation between
the two filters are negligible.

In the right panel of Fig. 8 we show how the 32 ETGs of our sample would be 
displaced at $z=0$ in the [$M_R$, R$_{e}$] plane in case of pure luminosity 
evolution, i.e. the R-band absolute magnitude plotted is 
$M_R(z=0)=M_R(z)-E(z)$.
In this case the different behaviour shown by the young ETGs with respect
to the old ETGs is even sharper than in the case of the KR (Fig. 4).
It is evident the agreement between the young ETGs and the 
local size-luminosity relation once considered their own luminosity evolution.
All the yETGs are located within the scatter region of the $z\sim0$ relation.
On the contrary, it is evident the disagreement between the old ETGs and 
the local S-L relation, disagreement which shows 
clearly that R$_e$ must change from $z\sim1.5$ to $z=0$.
It is worth noting that the calibration of the S-L relation
based only on the oETGs data is: 
$$log (R_e)=-0.28M_R(z=0)-5.6$$ (dashed (red) line in Fig. 8), a relation with 
the same slop of the one at $z\sim0$ but with an offset of 
$\sim0.5$ in the zero point corresponding to  a factor $\sim3$ 
in R$_e$.
 
In Fig. 9 our galaxies (filled symbols) are plotted on
the size-mass (S-M) plane and compared with the S-M relation found by 
Shen et al. (2003) for the local ETGs (solid line) expressed by the
following equation
 \begin{equation}
 R_e=2.88\times10^{-6}(\mathcal{M}_*/\mathcal{M_\odot})^{0.56}
 \end{equation}
The stellar mass they use is the one  from Kauffmann et al. (2003)
based on the Bruzual and Charlot (2003, BC03) models  and on
the Kroupa (2001) initial mass function while we used
the CB08 models and Chabrier IMF.
Longhetti et al. (2008) show that the stellar mass obtained with
 Kroupa IMF differs by less than 5\%  from the one obtained with Chabrier IMF
 and that the use of BC03 models leads to over-predict the mass estimate
 by  a factor 1.2-1.3 with respect to CB08 models.
 Thus, we decided not to apply any scaling factor to the relation found by 
 Shen et al. (2003) given such small differences.
Figure 9 shows that while 9 out of the 13  yETGs (70\%) follow
this size-mass relation at $z\sim0$ only 4 out of the 19 oETGs (20\%) agree 
with this relation.
The yETGs for which the luminosity evolution is requested to bring them
on the local S-L relation, do not need any mass or size evolution
since they naturally match the local S-M relation.
In other words, this is an evidence that yETGs are fully compatible
with a simple evolution from $z\sim1.5-2$ to $z=0$ of their 
$\mathcal{M}/L$ ratio due to pure luminosity evolution while their 
stellar mass remains unchanged.
On the contrary, the old ETGs follow well defined S-L and 
S-M relations but almost all of them have 
sizes much smaller than at $z\sim0$ implying that they
have changed significantly their structure from $z\sim1.5-2$ to $z=0$.
Using the relation of Shen et al. (e.q. 10) we have derived the
mean value $\langle f_{R_e} \rangle=1/N\sum(R_{e,0}/R_{e,z})$, 
i.e. the ratio between the radius R$_e$ of the 15 oETGs which do not follow 
the local S-M relation and the radius of the local ETGs with similar 
mass.
We have obtained $\langle f_{R_e} \rangle=2.6\pm0.5$
{ in agreement with the value 3.4$\pm1.7$ found by  Cimatti et al. (2008) 
in the same redshift range.}
Thus, the oETGs must increase their size by a factor 2.6 from $z\sim1.5$
to $z=0$, consistently with the result obtained from the KR and the
S-L relation.
It is worth noting that even hypothesizing that we have systematically 
overestimated by a factor 2 the mass of all the oETGs their
effective radius would be still 1.7 times larger then locally, i.e.
they would be 5 times denser.
To move the oETGs onto the local S-M relation their stellar mass
should be 6 times smaller, a factor too large to be accounted for
by any model assumption. 
\begin{figure}
\begin{center}
\hskip -0.2truecm
\includegraphics[width=9cm]{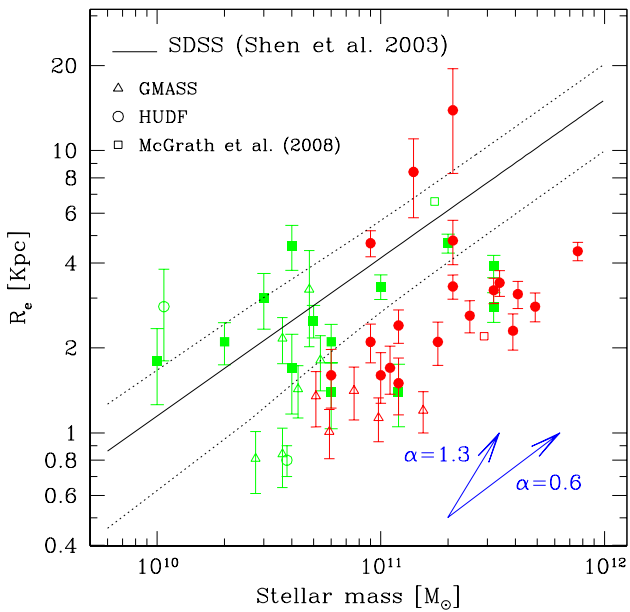} 
\vskip -0.5truecm
\caption{Size-mass relation for our sample of ETGs (filled symbols)
compared with the local relation found by Shen et al. (2003, solid line). 
The dotted lines represent the scatter of the relation.
Red symbols mark the oETGs while green symbols mark the yETGs
The two vectors represent the relation found by Boylan-Kolchin et al.
(2006) for the two extreme values $\alpha=0.6$ and $\alpha=1.3$.
The open symbols (triangles, circles, squares) indicate the galaxies of the 
GMASS sample (Cimatti et al. 2008), the two passive galaxies in the HUDF 
from  Daddi et al. (2005) and the two spheroids of McGrath et al. (2008).
We classified as old (and consequently marked with red color) those galaxies 
older than 2.0 Gyr.  
}
\end{center}
\end{figure}
{ In Fig. 9 a collection of ETGs at $1.3<z<2$ (open symbols) 
taken from the literature is also shown.
Triangles, circles and squares mark the GMASS galaxies from Cimatti et al.
(2008), the two HUDF galaxies from Daddi et al. (2005) and the two spheroids
of McGrath et al. (2008) respectively. 
The redshift and the spectral type of these ETGs are spectroscopically 
confirmed and the morphology is based on HST-ACS observations.
According to the analysis performed by the authors and to the parameters 
they derived, we divided this sample of ETGs in old and young defining old
those ETGs with age larger than 2 Gyr.
The old ETGs are marked by red open symbols while the young ETGs
are marked by green open symbols.
It is remarkable the agreement with the behaviour shown by our sample of
32 galaxies: young ETGs tend to distribute according to the local 
S-M relation (6 out of 9) while none of the old ETGs follow the local
S-M relation. 

It should be noted however that a non negligible fraction of yETGs, 
both in our sample of 32 ETGs and in the sample taken from the literature,
does not follow the S-M relation but follows the relation defined by the
old ETGs.
Thus, for some yETGs a major size evolution is still required.
If this result will be confirmed on a more solid statistical ground
proving that it is not due to an internal scatter in the estimate of the
physical parameters (age, stellar mass and effective radius), it implies 
that yETGs follow different histories of assembly and are
less homogeneous than old ETGs.
We will try to constrain their evolutionary path in the next section.}

\section{Constraining the formation and the evolution of ETGs}
The analysis performed in the previous sections shows that two populations
of ETGs exist at $z\sim1.0-2.0$.
They differ substantially for the age of their stellar populations
by about 2 Gyr and for the scaling relations they follow.
It is natural to ask how these two populations evolved from $z\lae2$ 
to $z\sim0$ to match the properties of the local ETGs 
and which is their assembly history they followed to 
have the properties shown at $z\sim1.5$.
We have tried to answer to these questions placing our results in 
the hierarchical paradigm of galaxy formation and evolution taking into 
account the results obtained from various renditions of merging models.

\subsection{Tracing the evolution at $z\lae 2$}
The older ETGs of our sample at $z\simeq1.5-2$ do not follow the S-M 
relation of local ETGs as well
as the other scaling relations.
Pure luminosity evolution from their redshift to $z=0$ does 
not bring them onto the local KR and S-L relation.
oETGs are characterized by effective radii R$_e\sim$2.5-3 times smaller than
those of the local ETGs with comparable surface brightness, 
absolute magnitude and stellar mass as deduced from the comparison with the
local KR, S-L and  S-M relations. 
Thus, an evolution of their size between $z\sim1.5-2$ and $z=0$ must occur 
to bring them onto the local scaling relations.
Such size evolution is often used to advocate the merging processes 
the ETGs
should experience during their life in the hierarchical paradigm of galaxy 
formation and evolution 
(e.g. Trujillo et al. 2004, 2007; Bell et al. 2006; De Lucia et al. 2006;
van Dokkum et al. 2008; van der Wel et al. 2008).
Merging is indeed usually invoked as the most obvious and efficient mechanism 
to increase the size of galaxies.

Boylan-Kolchin et al. (2006, BK06 hereafter), using simulations of 
dissipation-less merging, the so called "dry merging" (e.g. van Dokkum 
et al. 2005; Bell et al. 2006 and references therein) 
show that the remnants of dry mergers lie on the fundamental plane (FP,
Djorgovsky and Davis 1987; Dressler et al. 1987) of their progenitors.
However, the locations of the remnants in the projections of the FP, in
particular on the R$_e$-$\mathcal{M}_*$ relation, depends strongly  
on the merger orbit.
Thus, the projections of the FP can provide a tool to investigate
the assembly history of ETGs.
In their analysis, they find that the expected increase of the size of 
an ETG due to merging follows the relation 
R$_e\propto \mathcal{M}_*^\alpha$ with 
$0.6<\alpha<1.3$ (represented by vectors in Figure 9) 
depending on the orbital properties 
(see also Nipoti et al. 2002; Ciotti et al. 2007).
They show also that the index $\alpha$ is almost independent of the
mass ratio of the progenitors (see also Khochfar and Silk 2006 for 
a similar result) suggesting that their findings are applicable both 
to minor and major mergers. 
We have tried to consider this model of dry merging to increase the size of
oETGs.
We have seen that the effective radii of oETGs must increase by a factor 
$\langle f_{R_e} \rangle\simeq2.6$ from $z\sim1.5$ to $z=0$ in order to 
match the local S-M relation. 
Thus, the condition $R_f\simeq2.6R_i$ where 
$R_i$ and $R_f$ are the radii before (initial) and after (final) the merging
must be satisfied.
From the relation of BK06 it follows that
$\mathcal{M}_f^\alpha\simeq2.6\mathcal{M}_i^\alpha$ where $\mathcal{M}_i$ and
$\mathcal{M}_f$ are the masses before and after the merging.
Consequently the mass $\mathcal{M}_f$ that the
remnant must reach to increase the size 2.6 times is
\begin{equation}
\mathcal{M}_f=2.6^{1/\alpha}\mathcal{M}_i
\end{equation}
The most efficient way to move oETGs from their location onto the local 
S-M relation is for $\alpha=1.3$ (see vectors in Fig. 9), 
the maximum value found by BK06 which,
by the way,  minimizes the stellar mass of the remnant.
We thus obtain 
\begin{equation}
\mathcal{M}_f\ge2.1\mathcal{M}_i, \hskip1truecm \alpha\le1.3
\end{equation}
i.e. the mass of the remnant is at least twice the mass before the merging.
Any value of $\alpha$ lower than 1.3 would produce larger masses.
This result is difficult to reconcile with the number density 
of high-mass ETGs in the local universe.
Indeed, this mechanism would produce too 
much ETGs with masses much larger 
than 10$^{11}$ M$_\odot$ and an evolution in the stellar mass density at $z<2$
which is not observed (see e.g. Conselice et al. 2007).
For instance, Saracco et al. (2005) show that the 7 galaxies S2F1\# 
also studied in the present paper, account for 70\% of the local population 
of ETGs with comparable luminosity and mass.
If they twice their mass/luminosity, at $z\sim0$ we should observe 2-3 times
more ETGs with  masses $\mathcal{M}_*\ge4-5\times10^{11}$  M$_\odot$ then 
those in fact observed.   
Moreover, as previously noticed by Cimatti et al. (2008),
it is difficult to imagine that given all possible orbital
parameters in merging events, the effective value of $\alpha$ is
always close to the maximum one.
Finally, we recall that values $\alpha<1.3$ would worsen the
disagreement with the local number of high-mass ETGs.
Thus, we conclude that merging  
cannot be the mechanism with which oETGs increase their size 
at $z<2$ and that it is not the way to solve the problem.
Other mechanisms able to increase the size but to leave nearly unchanged the 
mass of ETGs must occur.
Close encounters or, more generally, interactions between galaxies can act 
in this way.
Their frequency and thus their efficiency depend on the number of close
encounters that a galaxy can experience in the last 9-10 Gyr of its life, 
a number that perhaps can be constrained from the statistics of pairs and 
from simulations.
Minor or "satellite" merging (e.g. Naab et al. 2007), i.e. merging between
galaxies with masses $\mathcal{M}_1$ and $\mathcal{M}_2$ in the ratio 
$\sim$0.1:1 or lower, would produce remnants with masses of the order of 
$\mathcal{M}_2$ but with larger size.
It is not clear the ability of this kind of merging in 
enlarging the size, however it could act in the right way
contributing to solve the problem of the small sizes of oETGs.

The younger population of ETGs, the yETGs, follows the size-mass relation 
of local ETGs, with few exceptions.
Luminosity evolution from $z\sim1.5-2$ to $z=0$, i.e. for fixed size R$_e$, 
would bring the yETGs onto the local KR and S-L
relation.
Thus, for these galaxies, the evolution of the $\mathcal{M}/L$ ratio
due to the expected luminosity evolution explains their 
observed properties at $z\sim1.5-2$ and brings them to agree with the 
scaling relations of local ETGs.
This suggests that the build-up of yETGs was already completed 
at $z\sim2$ providing no evidence in favor of merging 
at $z<2$ since it would bring them out of the S-M relation and the 
other scaling relations.
Indeed merging, to move yETGs along the S-M relation,
should take place for values $\alpha\simeq0.6$ (see vectors in Fig. 9)
producing remnants with masses $\mathcal{M}_f\sim5\mathcal{M}_i$.
For the analogous reasons discussed above, it is difficult to imagine 
that given all possible orbital parameters in merging events, the effective 
value of $\alpha$ for yETGs is always close to the minimum one.
Moreover, the luminosity at $z\sim0$ of the remnant should dim 
according to the increased size in order to match the KR and the S-L relation.
These requirements and fine tuning make this picture rather unlikely.
Finally, the reasoning relevant to the exceeding number of remnant ETGs 
with masses well in excess to $10^{11}$ M$_\odot$ applies also in this case,
and corroborates the conclusion that the assembly of yETGs was completed
at $z\sim2$ and that no merging has happened at $z<2$.
We cannot rule out that yETGs may experience satellite merging at $z<2$
if it leaves nearly unchanged their size besides their stellar mass.

For the reasons here discussed we can conclude that ETGs, both
young and old, have already reached their final stellar mass at $z\lae2$.
Major merging at redshift $z<2$, if any, must necessarily involve a 
negligible fraction of the old ETGs while satellite merging could involve 
both yETGs and oETGs even if in a different way. 

\subsection{Constraining the path at $z>2$: toward the formation of ETGs}
The older ETGs are characterized by a median age of about 
3.5 Gyr (and a dispersion of about 1 Gyr) which implies that their
stars formed at $z_f\sim5-6$.
Given the short time they have at disposal to form masses of
the order of $10^{11}$ M$_\odot$
of stars the star formation rate was necessarily $SFR>100$ M$_\odot$/yr 
(see also Cimatti et al. 2008, 2004; McCarthy et al. 2004; 
Daddi et al. 2005; Longhetti et al. 2005; Feulner et al. 2005; 
Kriek et al. 2006).
Moreover, oETGs are 2.5-3 times smaller than those at $z\sim0$
thus the physical mechanism(s) acting at $z>2$ must be capable 
also to produce very compact galaxies with stellar densities 15-30 times
higher than the local ones and than the yETGs.
Dissipational gas-rich merging 
can produce highly compact massive early-type galaxies 
if a high fraction of stars of the remnant formed during the merger
in a violent starburst (e.g. Kochfar et al. 2006; Springel and Hernquist 2005).
However, it is not clear whether the typical time scale of major merging 
 can fit with the requirements above ($\tau_{merge}>3$ Gyr from 
 Boylan-Kolchin et al. 2008).
Naab et al. (2007), with their {\em ab initio} hydrodynamics simulations, 
show that the early formation phase of galaxies can start with an initial
burst of star formation at $z\sim5$ accompanied
by mergers of gas-rich small sub-components and in situ intense 
star formation.
Although some fraction of the stars is accreted (the "quiescent" component,
Khochfar et al. 2006b) Naab et al. show that  this phase has the 
characteristics of a dissipative collapse since it happens on very
short timescale.
Thus, the compactness of ETGs produced through the gas-rich merging of
Kochfar et al. (2006) and the rapid dissipative collapse of Naab et al. (2007)
provide a scenario which seems to fit, at least qualitatively, the intense 
SFR required at $z\sim5-6$ and the need to assemble ETGs which at $z\lae2$
are compact and old.  

The younger population, the yETGs, is characterized by a median age of 
$\sim$1 Gyr and a dispersion of about 0.2 Gyr which push the last burst
of star formation at $z_f>2.5$.
In this case the constraints on the possible physical mechanism(s) acting
at $z>2$ are less stringent than those required to assemble oETGs at the 
same redshift.
A scenario in which yETGs accreted stellar mass by subsequent 
episodes of merging and by star formation in situ (which can also
be triggered by satellite and secondary merging events) can qualitatively
fit the properties  of these yETGs which  at $z<2$  must appear younger
and already enlarged in their size if compared with the local ones.
{ The few of them which appear much more compact than the others yETGs 
could be the results of the gas-rich merging scenario proposed by Kochfar 
et al. (2006) and Naab et al. (2007) characterized by a rapid dissipative 
collapse happened at $z\gae2.5$.}

From the observational point of view, the two populations of ETGs should
have different progenitors.
oETGs must experience a phase of intense star formation at high-z (z$\gae5$), 
yETGs can experience this phase at lower redshift or experience subsequent 
episodes of star formation possibly triggered by satellite merging.
{ In any case a different epoch of formation and assembly must characterize 
the two populations of ETGs.
This result qualitatively agrees with the model of Khochfar and Silk (2006)
which indicates that the scatter in the size of similar present-day
ellipticals is a result of their formation epoch, with smaller ellipticals 
formed earlier through mergers much richer of gas than the mergers
assembling larger ellipticals.
In Fig. 10 the predicted evolution of sizes for ETGs with
respect to the sizes of their local counterparts from the model 
of Khochfar and Silk (2006) is compared to the observed ratio 
R$_e(z)/$R$_e(z=0)$ for our sample of 32 ETGs.  
This ratio has been obtained for each galaxy dividing the observed 
effective radius  by that derived by the S-M relation of 
Shen et al. (2003) for the same mass of the galaxy.
As expected oETGs are preferentially located on the curve representing
the largest ratio which is expected for the  highest-mass ETGs since 
they should form much earlier in the model.
On the contrary yETG are preferentially located onto or even above 
the curves representing the  minimum size evolution which is expected for the 
lower mass ETGs which should form later.
It is worth noting that while this model reproduces very well the
observed relation between size evolution and formation epoch of ETGs
(the older the more compact/denser) the correlation with the stellar mass
is less evident from our data and a larger sample would be needed
to probe this issue. 
}
\begin{figure}
\begin{center}
\hskip -0.2truecm
\includegraphics[width=9cm]{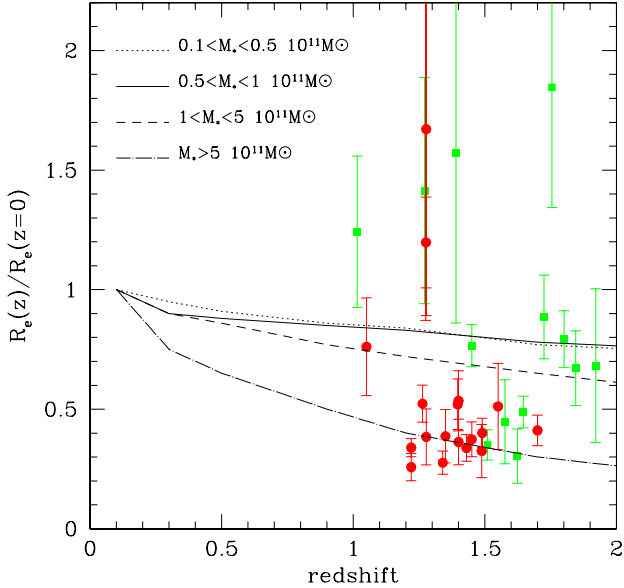} 
\vskip -0.5truecm
\caption{The ratio R$_e(z)/$R$_e(z=0)$ (filled symbols) obtained 
dividing the observed 
effective radius of our galaxies by that derived by the S-M relation 
at $z=0$ of 
Shen et al. (2003) for the same mass is compared to the predicted 
evolution of sizes (curves) from the model of Khochfar and Silk (2006).
Symbols are as in Fig. 8.
Dotted, solid, dashed and dot-dashed curves represent the predictions for
the different ranges of stellar masses listed in the legend.
}
\end{center}
\end{figure}
As to the progenitor candidates, as suggested by Cimatti et al. (2008), 
a possible population of progenitors could be the sub-mm selected galaxies 
seen at $z\gae3$ (e.g. Blain et al. 2002; Tacconi et al. 2006, 2008) whose
characteristics could fit those of ETGs at $1<z<2$. 
Even if it is difficult to identify the progenitors we can tray to
constrain their redshift.
Given the properties of oETGs (old and compact at $z\lae2$) and the typical 
time scale of merging ($\tau_{merge}>3$ Gyr), we should expect to see oETGs 
till $z\sim3-3.5$, 
the only  difference should be the age of their stellar population 
correspondingly younger.
On the contrary, yETGs should appear quite different at $z>2.5$,
most probably in the phase of merging, or star forming and interacting
with other galaxies.


\section{Summary and conclusions}
We presented the morphological analysis of a sample of 32 ETGs at $1<z<2$
with spectroscopic confirmation of their redshift and spectral type
based on HST-NICMOS observations in the F160W filter.
These 32 ETGs have been selected  from different samples and surveys 
on the basis of their spectroscopic and morphological
classification and are characterized by a multiwavelength coverage.
The HST-NICMOS observations in the F160W filter have allowed us
to derive the effective radius R$_e$ and the mean surface brightness 
$\langle\mu\rangle_e$ of galaxies in the rest-frame R-band of the galaxies,
less affected by morphological k-correction and star formation
then the optical bands usually used in the previous works.
Through the best fitting of their SEDs at known redshift we derived the 
R-band absolute magnitude, the stellar mass and the age for each of them.
The main results of the analysis we performed can be summarized as follows.

\begin{itemize}
\item
The 32 ETGs of our sample at $1<z<2$ are placed on the 
[$\langle\mu\rangle_e$,R$_e$] plane defining a relation with 
the same slope of the KR at $z\sim0$ but with a different
zero-point which accounts for the evolution they undergo
from $z\simeq1.5-2$ to $z=0$. 
We do not see differences between the 6 ETGs in clusters
and the other ETGs in agreement with other works (e.g. Rettura et al. 2008;
Gobat et al. 2008) even if the very low  statistics we have do not allow us
a detailed comparison of different environments.

\item 
The ETGs of our sample are composed of two distinct populations which
differ for the age of their stellar populations:
the older population, the oETGs, has a median age of about 3.5 Gyr
which implies that the bulk of stars formed at $z_f\sim5-6$;
the younger population, the yETGs, has a median age of about 1 Gyr and
correspondingly $z_f>2.5-3$.
Even if the absolute values of these ages can be model dependent, 
the different age
cannot be accounted for by any model assumption while it is reliable an age 
difference of about 1.5-2 Gyr for the two populations of ETGs.
yETGs tend to be less luminous and correspondingly less massive than oETGs
even if this tendency  is more
evident in the tails of the distributions. 

\item
yETGs follow the size-mass relation locally observed  (Shen et al. 2003)
and their expected luminosity evolution from $z\sim1.5-2$ to $z=0$ at 
fixed size R$_e$ brings them onto the local KR and size-luminosity relation.
Thus, their surface brightness and stellar mass density do not exceed those
of local ETGs with comparable luminosity and stellar mass, i.e.
no size evolution is required at $z<2$.
These properties suggest that young ETGs have already completed the growth of
their stellar mass at $z\sim2$ being them onto the S-M relation and that
they must evolve at $z<2$ purely in  luminosity to match the local KR and S-L
relation.
This provides  no evidence in favor of major merging  
at $z<2$ since it would bring them out of the local scaling relations.

\item
oETGs do not follow the local size-mass relation since they 
have sizes 2.5-3 times smaller then those provided by the S-M relation
at their stellar masses.
Pure luminosity evolution from their redshift to $z\sim0$ is not
sufficient to bring them onto the local KR and S-L relation.
Also in this case the effective radii are 2.5-3 times smaller than
the local ETGs with comparable surface brightness and absolute magnitude.
Thus, an evolution of their size at $z<2$ must occur to reconcile the oETGs
with the local population of ETGs.
Major (dry) merging at $z<2$ cannot solve the problem since it would produce
too much ETGs with stellar masses $>5\cdot10^{11}-10^{12}$ M$_\odot$ and
it should happens only under particular orbit conditions to move oETGs
onto the local S-M relation.
Other mechanisms able to increase the size and to keep constant the stellar
mass of oETGs (e.g. satellite merging, close encounters) must be invoked. 

\item
The different properties shown by the yETGs and the oETGs at $1<z<2$
imply  different evolutionary paths from their formation to $z\lae2$.
oETGs are much more compact and hence denser than local ones (15-30
times denser than the local ETGs with comparable stellar mass and than
yETGs) they are old with respect to the age of the universe at their redshift. 
Thus, their stellar mass must have formed at high-z ($z\sim5$) following a 
sort of dissipative gas-rich collapse ables to form rapidly most of the 
stellar mass thus producing a compact old remnant at $z\lae2$. 
This scenario is qualitatively fitted by the merging models of Khochfar
et al. (2006) and Naab et al. (2007).
Considering the typical time scale of merging ($\tau_{merge}>3$ Gyr; 
Boylan-Kolchin et al. 2008) and the age of the stellar population of oETGs,
we conclude that oETGs must exist as they are till $z\gae3-3.5$ with stellar 
populations correspondingly younger.
Progenitors should be searched for among the population of galaxies at
$z>3.5-4$ as high-mass galaxies with intense star formation. 
Possible candidates could be the sub-mm/selected galaxies  
(e.g. Blain et al. 2002; Tacconi et al. 2006, 2007)
as suggested by Cimatti et al. (2008).

The formation scenario for the yETGs seem to be rather different even if
less constrained with respect to the one of the oETGs.
The age of the yETGs  implies that the last burst of star formation  has 
taken place close to their redshift, i.e. $z\sim2.5$.    
They are not denser than the local one and are placed on the local S-M
relation.
They have completed their stellar mass growth as the oETGs but in a way such
that their size is larger. 
Major merging, satellite merging and close encounters coupled with star 
formation in situ can qualitatively fit these requirements producing  
at $z\lae2$ ETGs with
a young component of the stellar population and sizes comparable
to those of the local ETGs with similar stellar mass, surface brightness 
and luminosity.
This population of yETGs should appears at $z>2.5-3$ as star forming and/or 
interacting galaxies.
\end{itemize}
On the basis of the above results we believe that a key observational
test would be the measure of the velocity dispersion of oETGs and yETGs
since such quantity would unambiguously address the question whether
the two populations are dynamically different thus providing 
unique constraints on the mechanism of their formation and on their
size evolution at $z<2$.

\appendix
\section{Photometry of the sample}
In Table 1  the multiband photometry for
the whole sample of early-type galaxies is reported.
With the exception of the F160W-band magnitude which  we have estimated 
on the HST-NICMOS images, the other magnitudes have been taken from the
literature.

The  photometry of the two galaxies in the HUDF previously studied 
by Daddi et al. (2005) and by Cimatti et al. (2008) comes  
from the catalog of Coe et al. (2006). 
Namely, the B, V, R, z and J magnitudes are  the magnitudes measured
through the HST filters F435W, F606W, F775W , F850W and F110W respectively.
The Ks-band magnitude are taken from Cimatti et al. (2008) and comes
from VLT-ISAAC observations in the Ks filter.

For the sample of Longhetti et al. (2005), the magnitudes are those 
from the MUNICS survey  whose optical filters are slightly 
different from the standard Kron-Cousins filters.
A detailed description of the MUNICS photometry and of the filter
response is given by Drory et al. (2001).

The seven-filters (B, V, R, I, z, H, K) photometry of the 10 ETGs selected 
from the GDDS (Abraham et al. 2004) was originally taken from the photometric 
catalogs of the Las Campanas Infrared Survey (LCIR survey McCarthy et al. 2001).
The observations of the fields, the filters used and the photometric 
information are described in Chen et al. (2002) and Firth et al. (2002).
 
The photometry of the 6 galaxies at $z\sim1.27$ belonging to the cluster 
RDCS 0848+4453 in the Linx Supercluster
is taken from the paper of Stanford et al. (1997).
The three galaxies  CIG\_135, CIG\_108 and CIG\_142
seem to coincide with the three galaxies studied by van Dokkum et al. (2003).
The photometry is described in Stanford et al (1997) and subsequent 
observations are quoted in  Mei et al. (2006).
 
The photometry of the three galaxies in the HDF-N previously studied by 
Stanford et al. (2004) are partly taken from the HDF-N catalog (v2). 
The magnitudes in the filters B, V and I are in fact the magnitudes
measured through the HST filters F450W, F606W and F814W.

The photometry in the R, H and K bands of the galaxy 53W091 are taken 
from the papers of Dunlop et al. (1996) and Spinrad at al. (1997).
The I and J magnitudes are the magnitudes in the F814W and F110W filters
respectively taken from Waddington et al. (2002).

\section*{Acknowledgments}
We thank the anonymous referee for the useful comments and suggestions.
Based on observations made with the NASA/ESA Hubble Space Telescope, 
obtained at the Space Telescope Science Institute, which is operated by the 
Association of Universities for Research in Astronomy, Inc., under NASA 
contract NAS 5-26555. 
This work has received partial financial support from the Istituto Nazionale
di Astrofisica (Prin-INAF CRA2006 1.06.08.04) and from ASI 
(contract I/016/07/0).

\label{lastpage}

\end{document}